\documentclass[11pt]{article}
\usepackage{latexsym,amssymb}
\usepackage[centertags]{amsmath}
\usepackage{fullpage}
\title{
Adiabatic limit and the slow motion of vortices in a
Chern-Simons-Schr\"odinger system
}
\author{Sophia Demoulini and David Stuart
\\{\it{\small Centre for Mathematical Sciences, Wilberforce Road, 
Cambridge, CB3 OWB,
England}}\\{\it\small{ email:sd290@cam.ac.uk, dmas2@cam.ac.uk}}}
\date{} 
\begin{document}

\newtheorem{lemma}{Lemma}[subsection]
\newtheorem{theorem}[lemma]{Theorem}
\newtheorem{maintheorem}[lemma]{Main Theorem}
\newtheorem{corollary}[lemma]{Corollary}
\newtheorem{definition}[lemma]{Definition}
\newtheorem{remark}[lemma]{Remark}
\newtheorem{Notation}[lemma]{Notation}
\newcommand{\proof}{\noindent {\it Proof}\;\;\;}
\newcommand{\qed}{\protect~\protect\hfill $\Box$}

\newcommand{\be}{\begin{equation}}
\newcommand{\ee}{\end{equation}}
\newcommand{\ba}{\begin{eqnarray}}
\newcommand{\ea}{\end{eqnarray}}
\newcommand{\bes}{\[}
\newcommand{\ees}{\]}
\newcommand{\bas}{\begin{eqnarray*}}
\newcommand{\eas}{\end{eqnarray*}}
\newcommand{\hoa}{{H^{1}_{\mbA}}}
\newcommand{\hta}{{H^{2}_{\mbA}}}
\newcommand{\linf}{{L^\infty}}
\newcommand{\tp}{{\tilde P}}
\newcommand{\cf}{{\cal F}}
\newcommand{\ch}{{\cal H}}
\newcommand{\cfnd}{{\cal F}^{n}-{\cal F}^{n-1}}

\newcommand{\spsi}{_{{{\Psi}}}}
\newcommand{\pt}{\frac{\partial}{\partial t}}
\newcommand{\pxk}{\frac{\partial}{\partial {x^k}}}
\renewcommand{\theequation}{\arabic{equation}}
\newcommand{\rgt}{\rightarrow}
\newcommand{\lngrgt}{\longrightarrow}
\newcommand{\intsT}{ \int_{0}^{T}\!\!\int_{\Sigma} }
\newcommand{\dxdt}{\;dx\,dt}
\newcommand{\sublt}{_{L^2}}
\newcommand{\sublf}{_{L^4}}
\newcommand{\naf}{\nabla_\mbA \Phi}
\newcommand{\covt}{(\pt -iA_0) }
\newcommand{\ano}{A^{n}_{0}}
\newcommand{\aNo}{A^{N}_{0}}
\newcommand{\Psino}{\Psi^{n}_{0}}
\newcommand{\PsiNo}{\Psi^{N}_{0}}
\newcommand{\Nmo}{{N\!-\!1}}
\newcommand{\nmo}{{n-1}}
\newcommand{\nmt}{{n-2}}
\newcommand{\Nmt}{{N\!-\!2}}
\newcommand{\gotm}{\frac{\gamma}{2\mu}}
\newcommand{\ootm}{\frac{1}{2\mu}}
\newcommand{\tloc}{T_{loc}}
\newcommand{\tmax}{T_{max}}
\font\msym=msbm10
\def\Real{{\mathop{\hbox{\msym \char '122}}}}
\font\smallmsym=msbm7
\def\smr{{\mathop{\hbox{\smallmsym \char '122}}}}
\def\Complex{{\mathop{\hbox{\msym\char'103}}}}
\newcommand{\wkarr}{\; \rightharpoonup \;}
\def\Weak{\,\,\relbar\joinrel\rightharpoonup\,\,}
\newcommand{\To}{\longrightarrow}
\newcommand{\pa}{\partial_A}
\newcommand{\pbfa}{\partial_{\mathbf A}}
\newcommand{\pao}{\partial_{A_1}}
\newcommand{\pat}{\partial_{A_2}}
\newcommand{\dbar}{\bar{\partial}}
\newcommand{\barpa}{\bar{\partial}_{A}}
\newcommand{\barpbfa}{\bar{\partial}_{\mathbf A}}
\newcommand{\barpaphi}{\bar{\partial}_{A}\Phi}
\newcommand{\barpbfaphi}{\bar{\partial}_{\mathbf A}\Phi}
\newcommand{\myqed}{\hfill $\Box$}
\newcommand{\cd}{{\cal D}}
\newcommand{\dt}{\hbox{det}\,}
\newcommand{\sma}{_{{ A}}}
\newcommand{\bfpi}{{\mbox{\boldmath$\pi$}}}
\newcommand{\ce}{{\cal E}}
\newcommand{\ulh}{{\underline h}}
\newcommand{\ulg}{{\underline g}}
\newcommand{\ulX}{{\underline {\bf X}}}
\newcommand\la{\label}
\newcommand{\lamo}{\stackrel{\circ}{\lambda}}
\newcommand{\bfjo}{\underline{{\bf J}}}
\newcommand{\Vflato}{V^\flat_0}
\newcommand{\cm}{{\cal M}}
\newcommand{\dist}{{\mbox{dist}}}
\newcommand{\cs}{{\cal S}}
\newcommand{\mcF}{{\mathcal F}}
\newcommand{\mcn}{{\mathcal  V}}
\newcommand{\mce}{{\mathcal  E}}
\newcommand{\mcb}{{\mathcal B}}
\newcommand{\mca}{{\mathcal A}}
\newcommand{\mcdpsi}{{\mathcal D}_{\psi}}
\newcommand{\mcd}{{\mathcal D}}
\newcommand{\mcl}{{\mathcal L}}
\newcommand{\mclaphi}{{\mathcal L}_{(\mbA,\Phi)}}
\newcommand{\mcdadjp}{\mcdadj_{\psi}}
\newcommand{\mcdadj}{{\mathcal D}^{\ast}}
\newcommand{\zl}{{Z_\Lambda}}
\newcommand{\thl}{{\Theta_\Lambda}}
\newcommand{\ca}{{\cal A}}
\newcommand{\cb}{{\cal B}}
\newcommand{\cg}{{\cal G}}
\newcommand{\cu}{{\cal U}}
\newcommand{\co}{{\cal O}}
\newcommand{\smA}{\small A}
\newcommand{\ttheta}{\tilde\theta}
\newcommand{\tn}{{\tilde\|}}
\newcommand{\rbar}{\overline{r}}
\newcommand{\oeps}{\overline{\varepsilon}}
\newcommand{\cgl}{\hbox{Lie\,}{\cal G}}
\newcommand{\Ker}{\hbox{Ker\,}}
\newcommand{\const}{\hbox{const.\,}}
\newcommand{\Sym}{\hbox{Sym\,}}
\newcommand{\tr}{\hbox{tr\,}}
\newcommand{\grad}{\hbox{grad\,}}
\newcommand{\ttd}{{\tt d}}
\newcommand{\ttdel}{{\tt \delta}}
\newcommand{\ns}{\nabla_*}
\newcommand{\csl}{{\cal SL}}
\newcommand{\kr}{\hbox{Ker}}
\newcommand{\beq}{\begin{equation}}
\newcommand{\eeq}{\end{equation}}
\newcommand{\pr}{\hbox{proj\,}}
\newcommand{\proj}{{\mathbb P}}
\newcommand{\tproj}{\tilde{\mathbb P}}
\newcommand{\projq}{{\mathbb Q}}
\newcommand{\tprojq}{\tilde{\mathbb Q}}
\newcommand{\oN}{\overline N}
\newcommand{\cN}{\cal N}
\newcommand{\cmet}{{{\hbox{${\mathcal Met}$}}}}
\newcommand{\met}{{\hbox{Met}}}
\newcommand{\bfga}{{\mbox{\boldmath$\overline\gamma$}}}
\newcommand{\bfOmega}{\mbox{\boldmath$\Omega$}}
\newcommand{\bfTh}{\mbox{\boldmath$\Theta$}}
\newcommand{\bfmw}{{\bf m}_{\mbox{\boldmath$\smo$}}}
\newcommand{\bfmu}{{\mbox{\boldmath$\mu$}}}
\newcommand{\bfulX}{{\mbox{\boldmath${X}$}}}
\newcommand{\bfultX}{\tilde{\mbox{\boldmath${X}$}}}
\newcommand{\bfmuw}{{\mbox{\boldmath$\mu$}}_{\mbox{\boldmath$\smo$}}}
\newcommand{\dmug}{d\mu_g}
\newcommand{\ltwo}{{L^{2}}}
\newcommand{\lfour}{{L^{4}}}
\newcommand{\hone}{H^{1}}
\newcommand{\honea}{H^{1}_{\smA}}
\newcommand{\er}{e^{-2\rho}}
\newcommand{\onetwo}{\frac{1}{2}}
\newcommand{\lra}{\longrightarrow}
\newcommand{\dv}{\hbox{div\,}}
\newcommand{\mbA}{\mathbf A}
\def\smo{{\mbox{\tiny$\omega$}}}
\protect\renewcommand{\theequation}{\thesection.\arabic{equation}}

\font\msym=msbm10
\def\Real{{\mathop{\hbox{\msym \char '122}}}}
\def\R{\Real}
\def\Z{\mathbb Z}
\def\K{\mathbb K}
\def\J{\mathbb J}
\def\L{\mathbb L}
\def\D{\mathbb D}
\def\Mink{{\mathop{\hbox{\msym \char '115}}}}
\def\Integers{{\mathop{\hbox{\msym \char '132}}}}
\def\Complex{{\mathop{\hbox{\msym\char'103}}}}
\def\C{\Complex}
\font\smallmsym=msbm7
\def\smr{{\mathop{\hbox{\smallmsym \char '122}}}}
\def\ZHK{{\mathop{\hbox{\tiny ZHK}}}}

\maketitle
\thispagestyle{empty}
\vspace{-0.3in}
\begin{abstract}

We study a nonlinear system of partial differential 
equations in which
a complex field (the Higgs field) 
evolves according to a nonlinear Schr\"odinger equation,
coupled to an electromagnetic field whose time evolution is 
determined
by a Chern-Simons term in the action. In two space dimensions, 
the Chern-Simons dynamics
is a Galileo invariant evolution for $A$, which is an interesting 
alternative to the Lorentz invariant Maxwell evolution, and
is finding increasing numbers of applications in two dimensional
condensed matter field theory. The system we study, 
introduced by Manton, is a special
case (for constant external magnetic field, and a point
interaction) of the effective field theory
of Zhang, Hansson and Kivelson arising in studies of
the fractional quantum Hall effect. From the mathematical 
perspective the system is a natural gauge invariant generalization
of the nonlinear Schr\"odinger equation, 
which is also Galileo invariant and admits a self-dual structure
with a resulting large space of topological solitons
(the moduli space of self-dual 
Ginzburg-Landau vortices).
We prove a theorem describing the  adiabatic approximation  
of this system by a Hamiltonian system on the moduli space.
The approximation holds 
for values of the Higgs self-coupling constant  $\lambda$ 
close to the 
self-dual (Bogomolny) value of 1.  The viability of the 
approximation 
scheme depends
upon the fact that self-dual vortices form a symplectic 
submanifold of the 
phase space (modulo gauge invariance).  The theorem provides a 
rigorous 
description of slow vortex dynamics in the near self-dual limit.

\end{abstract}

\section{{Introduction and statement of results}}
\label{secint}
In this article we study vortex dynamics in a nonlinear
system  of evolution equations \eqref{scs}  introduced by Manton
(1997). This system is in fact a special case of an effective
field theory for the fractional quantum Hall effect (the 
Zhang-Hansson-Kivelson, or ZHK, model). In addition it is a natural
gauge invariant generalization of the nonlinear Schr\"odinger
equation, possessing important structural features (Galileo
invariance and self-dual structure with existence of 
related moduli spaces
of solitons) which make it interesting to study for
mathematical reasons. After introducing the system under study,
and putting it into mathematical and physical context, we
explain the necessary background material in order to
state our results, which appear in \S\ref{adlim}.

\subsection{Chern-Simons vortex dynamics}

We start by motivating the study of Manton's system from
the mathematical perspective, before going on to
show that it is equivalent to a special case of the ZHK 
model, and discussing its physical significance. 
\subsubsection{Manton's system on $\R^2$: mathematical context}

To introduce Manton's system, we start with the nonlinear 
Schr\"odinger equation on $\R^2$:
\beq
i\frac{\partial\Phi}{\partial t} = -\Delta\Phi
        -\frac{\lambda}{2}(1-|\Phi|^2)\Phi,
\la{nls1}\eeq
to be solved for $\Phi:\R\times\R^2\to\C$; $\lambda$ is a 
positive number.
This has the following properties:

(i)\;\;\; it defines a globally well-posed Cauchy problem,

(ii)\;\; it admits topological soliton solutions, 
the Ginzburg-Landau vortices, and 

(iii)\; it is invariant
under the group of Galilean transformations. 

Manton's system is a generalization of \eqref{nls1}, sharing
these properties, which
describes the evolution of 
a complex field $\Phi$, coupled
to a dynamically evolving electromagnetic potential 
$A=A_0dt+A_1 dx^1+A_2 dx^2$.
On $\R^2$ the system reads explicitly (writing 
$\langle a,b\rangle=\Re\bar a b$):
\begin{align} 
\begin{split}
  \frac{\partial A_1}{\partial t} + 
\frac{\partial }{\partial x^1} \bigl(
\frac{\partial A_2}{\partial x^1}-\frac{\partial A_1}{\partial x^2}
\bigr)\;=&\;
- \langle i\Phi ,\bigl(\frac{\partial}{\partial x^2}
-iA_2\bigr)\Phi\rangle\\ 
\frac{\partial A_2}{\partial t} + 
\frac{\partial }{\partial x^2} \bigl(
\frac{\partial A_2}{\partial x^1}
-\frac{\partial A_1}{\partial x^2}
\bigr)\;=&\;
+ \langle i\Phi ,\bigl(\frac{\partial}{\partial x^1}
-iA_1\bigr)\Phi\rangle
\\ 
 i\bigl(\frac{\partial}{\partial t}-iA_0\bigr) \Phi
+\sum_{j=1}^2\bigl(
\frac{\partial}{\partial x^j}-iA_j
\bigr)^2\Phi
\;=&\;
        -\frac{\lambda}{2}(1-|\Phi|^2)\Phi
\\
\frac{\partial A_2}{\partial x^1}-\frac{\partial A_1}{\partial x^2}
\;=&\;+ \frac{1}{2} (1-|\Phi|^2).
\end{split}\label{scsr2}
\end{align}
In addition to (i)-(iii) above, this system has the following
mathematical properties:

(iv)\; it is gauge invariant,

(v)\;\; self-dual structure and a 
large space of topological solitons (see \S\ref{mainstat}).

These properties make the study of vortex dynamics in Manton's
system interesting, since the self-dual structure makes a
rigorous analysis possible when the vortices are arbitrarily
close (see \S\ref{mainstat}-\ref{adlim}). The proof of our
results makes use of special mathematical features present
due to self-duality  which are
explained in \S\ref{idsd}; these features include
complex and symplectic structures
on the soliton moduli space, and a foliation of the phase
space which we call the Bogomolny foliation.

\subsubsection{Equivalence of Manton's system and a special case of 
the ZHK model}
\la{zhk}

The system \eqref{scsr2} can be derived from
the action $S=c_0\int s(A,\Phi) d^2x dt$, where
\beq
s(A,\Phi)=-\epsilon^{\mu\nu\rho}A_\mu\partial_\nu A_\rho
+\langle i\Phi,(\partial_t-iA_0)\Phi\rangle +A_0+
|\epsilon^{jk}\partial_j A_k|^2 + |(\partial_j-iA_j)\Phi|^2 
+\frac{\lambda}{4}(1-|\Phi|^2)^2
\notag\eeq
where Greek indices run over $\{0,1,2\}$ for space-time
tensorial quantities, Roman indices run over $\{1,2\}$,
and $\epsilon^{\mu\nu\rho}, \epsilon^{jk}$ 
are the 
completely anti-symmetric symbols and the summation convention
is understood. This action is one of a 
class involving
the Chern-Simons term $\epsilon^{\mu\nu\rho}A_\mu\partial_\nu A_\rho$,
see \cite{gd,hz} for a review. 
It is characteristic of these theories
that variation of the action with respect to $A_0$ gives a
constraint equation involving the magnetic field, in this 
case the final equation of \eqref{scsr2}. This equation is 
analogous to the Gauss law in ordinary Maxwell theory, and is
referred
to as a constraint because the previous (dynamical) equations in 
\eqref{scsr2} imply that its time derivative
vanishes (exactly as do the dynamical 
Maxwell equations for the Gauss law). This constraint means
that many apparently different actions give rise to the same
Euler-Lagrange equations: in particular we can replace the
above action density with
\beq
\tilde s(A,\Phi)=-\epsilon^{\mu\nu\rho}A_\mu\partial_\nu A_\rho
+\langle i\Phi,(\partial_t-iA_0)\Phi\rangle +A_0+
 |(\nabla-i\mbA)\Phi|^2 
+\frac{\lambda+1}{4}(1-|\Phi|^2)^2.
\notag\eeq

We now introduce the ZHK
action $S_{\ZHK}(a,\Phi;A^{ext})=c\int s_{\ZHK} d^2x dt$
and
show that Manton's action $S$ is in fact a special case of
$S_{\ZHK}$; essentially the same observation appears also 
in \cite[page 54]{hz} and \cite[Section 4.8]{rt02}. 
The ZHK action is the action for a mean field description of the
quantum Hall effect. This effect refers to the current 
$J_j=\sigma_{jk}E^{ext}_k$ produced in an effectively 
two dimensional system of
electrons in a {\em strong} transverse magnetic field, by application
of an applied electric field $E^{ext}_k$. In the right experimental
situation the conductivity tensor $\sigma_{jk}$ is found to be 
off-diagonal (i.e. $\sigma_{11}=0=\sigma_{22}$), with the non-zero
entries $\sigma_{12}=-\sigma_{21}=f e^2/\hbar$, where $f$ is an 
integer, or a fraction, for (respectively) the integer and 
fractional quantum Hall effect. This quantization of the values of
$\sigma_{12}$ means that as the number of charge carriers is increased
there is no corresponding increase in the current - it lies
on a plateau - at least until
the number of carriers is sufficiently greatly increased, 
at which point the conductivity
moves to another of the quantized values,
and the current moves to another plateau.
In the mean field description 
the field $\Phi$ interacts with an external (applied)
electromagnetic potential $A^{ext}$ and a ``statistical''
potential $a$, according to:
\begin{align}
s_{\ZHK}=\frac{\kappa}{2}\epsilon^{\mu\nu\rho}
a_\mu\partial_\nu a_\rho
&+\langle i\Phi,(\partial_t-ia_0-iA^{ext}_0)\Phi\rangle+
\frac{1}{2m} |(\nabla-i\mathbf{a}-i\mbA^{ext})\Phi|^2 \notag\\
&+\int(1-|\Phi(x)|^2)V(x-x')(1-|\Phi(x')|^2)d^2x',
\notag
\end{align}
(see \cite{zhang92}, \cite[Section 4.6]{gd}, or \cite[Equations (7)-(8)]{zhk},
taking note of the published erratum for the latter reference).
To reduce this to $\tilde s$ we consider the case of 
a constant external magnetic field 
$B^{ext}=\partial_1 A^{ext}_2
-\partial_2 A^{ext}_1$ with $A_0^{ext}=0$. (The standard 
configuration in quantum Hall experiments involves a {\em strong}
 transverse magnetic field applied to an effectively two 
dimensional electron gas, with relatively small electric
potentials applied along one of the planar directions.)
Define
$A=a+A^{ext}$. Now check that
\begin{align}
\epsilon^{\mu\nu\rho}a_\mu\partial_\nu a_\rho
&=a_0(\partial_1 a_2-\partial_2 a_1)-a_1(\partial_t a_2-\partial_2 a_0)+a_2(\partial_t a_1-\partial_1 a_0)\notag\\
&=A_0(\partial_1 A_2-\partial_2 A_1-B^{ext})
-(A_1-A_1^{ext})(\partial_t A_2-\partial_2 A_0)\notag\\
&\qquad\qquad\qquad\qquad\qquad\qquad\qquad\qquad
+(A_2-A_2^{ext})(\partial_t A_1-\partial_1 A_0)\notag\\
&=\epsilon^{\mu\nu\rho}A_\mu\partial_\nu A_\rho
-2A_0 B^{ext}+\partial_t(A_1^{ext}A_2-A_2^{ext}A_1)
+\partial_1(A_0 A_2^{ext})-\partial_2(A_0 A_1^{ext})
\notag\\
&=\epsilon^{\mu\nu\rho}A_\mu\partial_\nu A_\rho
-2A_0 B^{ext}+
\epsilon^{\mu\nu\rho}\partial_\mu(A^{ext}_\nu A_\rho)\notag
\end{align}
and deduce that $s_{\ZHK}-\tilde s_{\ZHK}$ is a derivative,
where
\begin{align}
\tilde s_{\ZHK}=\frac{\kappa}{2}\epsilon^{\mu\nu\rho}
A_\mu\partial_\nu A_\rho
&-\kappa B^{ext} A_0
+\langle i\Phi,(\partial_t-iA_0)\Phi\rangle+
\frac{1}{2m} |(\nabla-i\mbA)\Phi|^2 \notag\\
&+\int(1-|\Phi(x)|^2)V(x-x')(1-|\Phi(x')|^2)d^2x'.
\la{zhka}
\end{align}
Now recall that derivatives in the action density 
do not affect the corresponding Euler-Lagrange equations (they are
null Lagrangians). It follows by comparing 
$\tilde s$ and $\tilde s_{\ZHK}$ that the equations of motion
for $S_{\ZHK}$ will be identical to those of Manton if we choose
$V(x)=(\lambda+1)\delta(x)/4$, $\kappa=-2$,$B^{ext}=1/2$ and
$m=1/2$.
Therefore, we conclude that,
at least as far as the classical equations of motion 
are concerned,
the ZHK model with these values 
is the same as Manton's
system in the case of
\begin{itemize}
\item
a constant external magnetic field of appropriate value, and
\item a point interaction $V(x)\propto\delta(x)$.
\end{itemize}

We now discuss the physical interpretation of the model in the 
fractional quantum
Hall context. There is a microscopic model,
due to Laughlin, which explains the observed phenomena in a 
well-accepted way in terms of a new phase of the two
dimensional electron gas (for low temperature and high
magnetic fields), with ground state described by the
Laughlin wave function. There is an energy gap in the spectrum,
so that the excitations above this new ground state have
strictly positive energy - the Laughlin quasi-particles
and quasi-holes, which have fractional statistics 
and fractional charge.
It is this fractional charge, combined with the explanation
of the integer quantum Hall effect, which gives rise to the 
fractional quantum Hall effect.
The effective field theory
proposed in \cite{zhk}, and reviewed at length
in \cite{zhang92}, gives a mean field description which is
not expected to be accurate on microscopic length scales,
but which does give an alternative explanation of all the main
observed phenomena. In this mean field theory, the
elementary excitations are described by the topological 
vortices, which
are endowed with the same fractional charge and 
fractional statistics as the Laughlin quasi-particles.
(The Chern-Simons term for the statistical gauge field
$a$ in the action serves to change the
statistics in the well known way explained in 
\cite{zhang92,aswz}).
It is understood that, in the mean field picture,
it is the pinning of vortices which explains the 
observed plateaus in the Hall conductance (\cite{pg,sto90,zhang92}). 
Thus a good understanding
of vortex dynamics in the ZHK model should be
useful to gain a better explanation of the phenomena within
the context of the mean field approach. Needless to say
there is still much work to be done to go from the results
of this paper, to results which would apply directly to
the experimental situation: even apart from issues like
the spatial domain and
the real values of the coefficients in the model,
it will be necessary to treat the applied electric 
potential which produces the Hall current flow. This means
that in the above derivation we should allow for an external
electric field $E^{ext}_k=-\partial_k A_0$, in addition to the
static magnetic field, and investigate its effect on the
vortex motion. Since the magnetic field is very strong in the
experimental situation, it should be reasonable to treat the
electric field perturbatively.

To conclude the discussion of the motivation for our work, the system
\eqref{scsr2} is one of a class of dynamical Chern-Simons vortex
models whose study is mathematically interesting (due to properties
(i)-(v) above), and which is physically relevant (as we have just discussed).
The use
of such models in condensed matter applications is
phenomenological, so the precise Lagrangian and many values of
the coupling constants, etc. are not precisely known. 
(Actually, 
in  \cite{zhk,zhang92} 
the ZHK action \eqref{zhka} is derived formally
from an ostensibly microscopic, second quantized, action. However, 
this microscopic action itself seems to have
a phenomenological character, since it involves
excitations which are not fundamental electrons, but rather
collective excitations - see the dicussion following
(2.6) in \cite{sto90}).
In any case
our main result does
provide a rigorous basis for understanding vortex dynamics in a
prototype for a class of theories which are of interest 
in two dimensional
condensed matter theory. The adiabatic limit system \eqref{exh}
which we derive for the vortex dynamics
cannot usually be written down explicitly, but 
as discussed in remark \ref{phenrem},
the behaviour of
some of its solutions can be understood
in many cases, and thus information on
the dynamics of vortices can be deduced within the 
framework of this approximation. There are reasons
to hope that qualitative features of the motion
in this limiting situation will have a wider validity: 
see remark \ref{phenrem}.

As a final comment on the quantum Hall effect, 
there is another type of soliton - 
a nonlocal Skyrmion - which appears in treatments of the
ferromagnetic properties of quantum Hall samples
(see \cite{skkr,ts03,g99} 
and \cite{sdds07} for some analytic properties
of these Skyrmions in a particular case).

\subsubsection{General physical context for Chern-Simons models}

There has been a fairly long standing interest in systems of the
type \eqref{scsr2} in the physics literature;
we give a brief summary and  refer the reader to \cite{jp,gd,hz}
for detailed reviews.
The study of Chern-Simons dynamics in 2+1 dimensional Maxwell and
(non-abelian gauge) theories was started in the early 80's (see e.g.
\cite{djt}) and the incorporation of vortices into this dynamics (in
systems with coupling to a nonlinear Schr\"odinger equation) has been
studied since at least the early 90's by theoretical physicists (see
the review \cite{jp} for early work on Chern-Simons vortices). 
The reason
for this interest is both because (i) the Chern-Simons models are used
widely in condensed matter physics in descriptions of
the quantum Hall effect and high T
superconductivity, and (ii) because they provide a useful scenario in
which to probe certain complex issues in field theories.

Regarding the first point, there are various time-dependent models
for magnetic vortices but at very low temperatures it is argued
(\cite{atz,sto}) that the motion should be non-dissipative so the usual
Eliashberg-Gorkov equation is not appropriate, and the Chern-Simons
coupled to Schr\"odinger 
vortex dynamics is widely used instead in the condensed matter
literature, both in superconductivity and the quantum Hall effect;
see \cite[Section 10.7]{pg}, \cite[Chapter 6]{n} for general
discussions, in addition to the references for the ZHK model
in the previous section.
(Relativistic invariance is broken in these condensed matter
applications, so the corresponding relativistic abelian Higgs model,
whose vortex dynamics are
studied in \cite{s94}, is not appropriate. The main application
which has been suggested for the relativistic dynamics 
appears to be cosmic string evolution.) There have been 
explanations offered for the  wide occurence of Cherns-Simons
types models in two dimensional condensed matter applications 
in terms
of universality features of large scale effective 
actions for two dimensional interacting
electronic and magnetic systems with spin (\cite[Section 3]{fs92}).

Regarding the second reason for interest in these models,
it was realized in the 1980's that
in two dimensions there were possible quantum statistics
other than the usual fermionic and bosonic types - anyons,
are two dimensional quantum particles undergoing an arbitrary
phase shift on interchange. Furthermore, composite objects
made up from charged particles orbiting vortices (or flux tubes)
have fractional spin and statistics (\cite{wilc82}).
In \cite{frm} the authors
study the quantum theory of a Lagrangian which is closely related
to \eqref{mantlag}, and use it to investigate the quantization of
solitons, quantum statistics and anyons in a 
rigorous quantum field theory setting. 

\subsection{Organization of the article}
The article is organized as 
follows. Our main aim is the study of vortex dynamics in the 
Chern-Simons-Schr\"odinger system with spatial domain a Riemann surface, so
we start in the next section by writing down
the equations in this case, 
and then giving necessary background including
a discussion of the self-dual vortices in 
\S\ref{mainstat}. We
then state our main result, theorem \ref{main}, which
describes the adiabatic approximation of vortex motion in the
self-dual limit. This is proved in \S\ref{pmainth} 
following a strategy explained in the context of a 
simple model problem in \S\ref{simple}. The proof
uses some specialized identities related to the
self-dual (or Bogomolny) structure,  presented 
in \S\ref{idsd} (which may be read separately). Various
subsidiary facts and lemmas are given in the appendix.

\subsection{The equations on a surface}
\la{eq}

The dependent variables are a complex field $\Phi(t,x)$,
and an electromagnetic potential 1-form
$$
A_0dt+A_1dx^1+A_2 dx^2.
$$
This 1-form determines a covariant derivative operator
\beq
D=(D_0,D_1,D_2)=(\pt-iA_0,D_1,D_2)=(\pt-iA_0,\nabla_1-iA_1,\nabla_2-i A_2),
\la{covd}
\eeq
which in turn determines the
electric field $E=E_jdx^j$ and magnetic field
$B(t,x)$ via \eqref{curv}; all these fields are
defined for $(t,x)\in\R\times\Sigma$ where
$\Sigma$ is a two dimensional spatial domain, 
taken to be a Riemann surface with
metric $g_{jk}dx^jdx^k$, area form $d\mu_g$
and complex structure 
$J:T^*\Sigma\to T^*\Sigma$ 
(where $j,k,\dots$ take values in $\{1,2\}$ and 
we use the summation convention). Introducing a covariant Laplacian 
operator by 
$$
-\Delta_A\Phi=-\frac{1}{\sqrt{\dt  g}}
D_j\bigl(g^{ij}\sqrt{\dt g}D_i\Phi\bigr)
$$
(using a local frame and coordinates),
the equations are
\begin{align} 
\begin{split}
  & E_j + \frac{\partial B}{\partial x^j} =
- J_j^k\langle i\Phi ,{D}_k\Phi\rangle\\ 
  & i(\pt-iA_0) \Phi = -\Delta_A\Phi
        -\frac{\lambda}{2}(1-|\Phi|^2)\Phi
\\
& B = \frac{1}{2} (1-|\Phi|^2).
\end{split}\label{scs}
\end{align}
The electric and magnetic field can be combined to
give the space-time electromagnetic field
$$F_{\mu\nu}dx^\mu\wedge dx^\nu=E_j dt\wedge dx^j
+Bd\mu_g.$$ This two form is obtained as the commutator
of the space-time covariant derivative \eqref{covd}
which mediates the coupling in \eqref{scs}:
\beq\la{curv}
[D_{\mu},D_\nu]\Phi=-iF_{\mu\nu}\Phi,\quad\hbox{where}
\; F_{0k}=E_k,
\;\hbox{and}\; \frac{1}{2}F_{jk}dx^j dx^k=Bd\mu_g.
\eeq
(Greek indices run through $0,1,2$ and Latin indices 
through $1,2$ only. Bold face is used to indicate the spatial
part of a vector or one-form etc., except in 
\S\ref{idsd} where time does not appear at all.)

We now describe this set-up briefly
in geometrical terms. Assume given a
one dimensional complex vector bundle
$L\to\Sigma$, with
a real inner product $h$ locally of the form $\langle a,b\rangle=h\Re\bar ab$,
and corresponding norm $|a|^2=\langle a,a\rangle$; if we employ a unitary
frame over some chart then  $\langle a,b\rangle=\Re\bar ab$.
We are
then solving for
an $S^1$ connection 
on the
bundle ${\L}\equiv\R\times L\to\R\times\Sigma$,
with associated covariant derivative $D$,
and a section $\Phi$ of ${\mathbb L}$. 
To be more explicit, fix a smooth connection on $L$ determined
by a covariant derivative operator $\nabla$, so that the
spatial part of $D$, which will be written
${\bf D}$,  takes the form $D_j=\nabla_j-iA_j$
for a real 1-form ${\mathbf A}=A_jdx^j \in \Omega^1_\R(\Sigma)$; here
$\nabla$ is independent of time. (It is generally not possible
to choose $\nabla$ to be flat, and it will have a curvature,
determined by a function $b$ such that 
$[\nabla_j,\nabla_k]\Phi dx^jdx^k=
-ibd\mu_g\Phi$; it is always possible to choose $b=\const$,
and we will do this throughout.) In any case, with this procedure
the space of connections on $L$ can be identified with the space of
real one-forms.
Then at each time $t\in\R$ 
we are solving for a section
$\Phi(t)$ of $L$, a 1-form 
${\mathbf A}(t)=A_1(t)dx^1+A_2(t)dx^2$ on $\Sigma$, and a 
real valued function $A_0(t)$ on $\Sigma$. 
The electric field is given by 
$$E_j=\frac{\partial A_j}{\partial t}-\frac{\partial A_0}{\partial x^j} $$
and the magnetic field by
$$
Bd\mu_g=bd\mu_g+{\mathbf{d}\mbA}.
$$
(Here, and elsewhere, 
we write ${\mathbf d}$ in bold face when it is necessary to 
indicate that only the spatial part is taken.)
The 2-form
$-i E_jdt\wedge dx^j-iBd\mu_g$ is the 
curvature associated to
the space-time covariant derivative
$D$, as in \eqref{curv}.
For the case $\Sigma=\Real^2$, the system was proposed by 
Manton (1997), who derived it 
as the Euler-Lagrange equation for the
Lagrangian \eqref{mantlag}.

\begin{Notation}\la{note}
We shall always consider conformal co-ordinate systems on $\Sigma$ in which the metric is
of the form $g=e^{2\rho}\bigl((dx^1)^2+(dx^2)^2\bigr)$ and the volume element is then
$e^{2\rho}dx^1\wedge dx^2$. On functions the Hodge operator acts as
$*f=f d\mu_g=fe^{2\rho} dx^1\wedge dx^2$ and $*^2=1$, so that
$*d\omega=e^{-2\rho}(\frac{\partial \omega_2}{\partial x^1}
-\frac{\partial \omega_1}{\partial x^2})$ for 1-forms $\omega$.
On 1-forms $*(\omega_1dx^1+\omega_2dx^2)=
\omega_1dx^2-\omega_2dx^1$, which
is just the negative of the
complex structure $J$, 
represented in conformal co-ordinates 
by the anti-symmetric tensor
$J^j_i$ with $J^1_2=-1,J^2_1=+1$, the other components being zero.
Correspondingly we decompose a one-form as $\omega=
\omega^{(1,0)}dz+\omega^{(0,1)}d\bar z$; in particular for the derivative
$df=\partial f dz+\bar\partial fd\bar z$, 
with $\bar\partial f=\frac{1}{2}
(\frac{\partial f}{\partial x^1}+
i\frac{\partial f}{\partial x^2})$,
and
$$
{\bf D}\Phi=
{D}^{(1,0)}\Phi+D^{(0,1)}\Phi
=\partial_{\mathbf A}\Phi dz+\barpbfaphi d\bar z,
$$
with $\barpbfaphi=\frac{1}{2}\Bigl((\nabla_1-iA_1)+i(\nabla_2-iA_2)\Bigr)\Phi$ 
etc.; see \S\ref{idsd}.
For a 1-form ${\mathbf A}$ we write the co-differential 
${\mathbf d}^*{\mathbf A}=-\dv {\mathbf A}$, with 
$\dv {\mathbf A}=e^{-2\rho}(\frac{\partial A_1}{\partial x^1}
+\frac{\partial A_2}{\partial x^2})$, and
the Laplacian on real functions is 
$\Delta f=e^{-2\rho}\frac{\partial^2 f}{\partial x^i\partial x^i}$, 
(with the
summation convention), and on sections of $L$ the covariant Laplacian is 
$-\Delta_{\mathbf A}\Phi=e^{-2\rho}(D_1^2+D_2^2)\Phi$ when a unitary
frame is used. The operators
$\dv, *d,\Delta$ (resp. $\Delta_{\mathbf A}$) 
all depend on $g$ (resp. $g,h$), but this
is not indicated as $g,h$ are fixed, and similarly dependence of constants
in estimates on $(\Sigma,g)$ and $h$ will be suppressed  
throughout the article.
\end{Notation}
\begin{Notation} We are dealing with sections of
smooth vector bundles $V$ over $\Sigma$ with an inner product
$\langle\cdot,\cdot\rangle$
induced from the Riemannian metric $g$ and the metric $h$ on
$L$ in the standard way; since $g,h$ are fixed throughout 
they will not be indicated. Thus, for example,
$$
|{\bf D}\Phi|^2=e^{-2\rho}\Bigl(\langle {D_1\Phi},D_1\Phi\rangle+
\langle D_2\Phi,D_2\Phi\rangle\Bigr).
$$
We write $\Omega^0(V)$ for the smooth sections of $V$
and $\Omega^p(V)$ for the smooth $p$-forms taking values 
in $V$. 
We will make use of the
Sobolev spaces $H^s(V)$ of sections of $V$ whose coefficient functions
(in any frame over any open set $\Omega\subset\Sigma$)
lie in the standard Sobolev space $H^s(\Omega)$; the
corresponding Sobolev space of $V$-valued 
$p$-forms is denoted $H^s(\Omega^p(V))$. In \S\ref{secint}
and \S\ref{pmainth}
we shall generally omit explicit reference to the vector
bundle, since this is usually clear, and write $H^s$ in place
of $H^s(V)$ etc.
(and $\|\,\cdot\,\|_{H^s}$ for the corresponding norms). However
if it is necessary to emphasize that time is fixed, and the
norm is taken over $\Sigma$, we shall write $H^s(\Sigma)$. 
\end{Notation}

Further 
notational conventions are given in the appendix and in 
\S\ref{idsd}, particularly in relation to the complex
structure (see also the textbook \cite[\S 9.1]{jost98} for
a treatment of the background material).

\subsection{Existence theory for the Cauchy problem}
\la{exsec}

Inherent to the system \eqref{scs} is 
the property of {\it gauge invariance}:
let $\chi(t,x)$ be a smooth real valued function, then $(A,\Phi)$ is a
smooth solution if and only if $(d\chi+A,\Phi e^{i\chi})$ is. This
introduces a large degeneracy to the solution space which may be
removed by a choice of gauge in various ways. 
We will adopt here the following
gauge condition which involves the time derivatives $\dot \mbA,
\dot\Phi$, of $\mbA,\Phi$:
\begin{equation}
\label{gauge1}
\dv \dot \mbA -
\langle i\Phi,\dot\Phi\rangle\equiv 
e^{-2\rho} (\partial_1 \dot{A}_1 + \partial_2 \dot{A}_2) -
\langle i\Phi,\dot\Phi\rangle = 0.
\end{equation}
We make this choice because it allows a convenient
description of the complex and symplectic structures 
on the moduli space of vortices (see remark \ref{bogom}
and \S\ref{idsd}), and also is useful in the derivation
of energy estimates for the time derivatives (see
\S\ref{gl} and \S\ref{pme}).
In this gauge  global existence can be stated as follows:

\begin{theorem}[Global existence in gauge \eqref{gauge1}]
\label{thglob2} 
Consider the Cauchy problem for \eqref{scs} with initial data
$\Phi(0)\in H^2 (\Sigma)$ and  ${\bf A}(0) \in H^1(\Sigma)$.
There exists a global
solution satisfying \eqref{gauge1} and the estimate 
\begin{equation}
|\Phi(t)|_{H^2(\Sigma)}\le ce^{\alpha e^{\beta t}}
\label{globest}
\end{equation}
for some positive  constants $c, \alpha , \beta$ depending only on
$(\Sigma,g)$, the equations, and the initial data.
The solution has regularity
$\Phi\in  C\bigl([0,\infty); H^2 (\Sigma)\bigr)
\cap C^1\bigl([0,\infty); L^2 (\Sigma)\bigr)$
and $\mbA\in  C^1\bigl([0,\infty); H^1 (\Sigma)\bigr)$.
If the initial data are smooth, then the solution
is also smooth.
\end{theorem}

It is explained in appendix \ref{brez} how to derive this
theorem from the global existence result of \cite{ds07}, which
is stated in another gauge.
Bounds of the type \eqref{globest}
were derived in \cite{bg} for the cubic nonlinear 
Schr\"odinger equation on $\R^2$, by means of the inequality
\begin{equation}
|u|_{L^\infty}\le C[1+\sqrt{\ln(1+\|u\|_{H^2})}],
\end{equation}
valid for $u\in H^2 (\R^2)$ and with $C=C(\|u\|_{H^1})$. The
proof of global regularity for \eqref{scs} depends on a covariant
version of this inequality (given in lemma \ref{covbg}), and
a careful treatment of various commutator terms $[D_\mu,D_\nu]$
which indicates that they have a comparable strength 
to the cubic nonlinear term.

In conclusion, theorem \ref{thglob2} provides
a global solution which is a continuous curve in the
space $\ch_2$ where for $s\in\R$ we define
\beq
\la{h2}
\ch_s\equiv\{
(\mbA,\Phi)\in H^{s-1}(\Sigma)\times H^s(\Sigma)
\},
\eeq
with the corresponding norm $\|\,\cdot\,\|_{\ch_s}$.
From now on we will consider only $({\mathbf A},\Phi)$ which lie
(at a given time) in the space $\ch_2$. The gauge group
at fixed time is given by
\beq
\la{gg}
\cg\equiv\{g\in H^2(\Sigma; S^1)\}
\eeq
and acts on $\ch_2$ according to 
$g\cdot(\mbA,\Phi)=(\mbA+g^{-1}dg,\Phi g)$.
(Restricting to the set where $\Phi$ is not identically zero the action
is free and gives a principal $\cg-$bundle structure. The gauge condition
\eqref{gauge1} can be then regarded as giving a connection - i.e. a
family of horizontal subspaces -  on this bundle.)

\subsection{Variational and Hamiltonian formulation}
\la{varham}

The equations \eqref{scs} can be derived formally as the 
Euler-Lagrange equations associated to the functional
\beq
S(A,\Phi)=\frac{1}{2}\int_{\smr\times\Sigma} -A\wedge F
+\bigl(\langle i\Phi,D_0\Phi\rangle +A_0
 +2v_{\lambda}(A,\Phi) \bigr)dt d\mu_g,
\la{mantlag}
\eeq
where
\begin{equation}
{v}_{\lambda}(\mbA, \Phi)  =
\frac{1}{2}\Bigl(B^2 + |{\bf D}\Phi|^2 
+\frac{\lambda}{4}(1-|\Phi|^2)^2 \Bigr)
\label{glen}\end{equation}
is the density of  the Ginzburg-Landau static energy. (The parameter 
$\lambda$ is a positive real numbers).
Although $S$ is not manifestly gauge invariant it changes by an exact form under gauge 
transformation, and the Euler-Lagrange equations \eqref{scs} are gauge invariant.
Vortices are critical points of the static energy 
$$\mcn_{\lambda}(\mbA,\Phi)
=\int_\Sigma v_{\lambda}(\mbA,\Phi)d\mu_g,$$ 
as will be discussed further in the next
section.

To see that the system
\eqref{scs} is Hamiltonian, 
observe that there is a complex structure on the phase space $\ch_2$ given
by $\J:(\dot  \mbA,\dot \Phi)=(-J\dot  \mbA,i\dot \Phi)$ which allows the
introduction of a symplectic structure $\Omega(v,w)=\langle\J
v,w\rangle$ where $\langle\,\cdot\, ,\,\cdot\,\rangle$ is the $L^2$
inner product. Using this symplectic form the system \eqref{scs},
in temporal gauge $A_0=0$, is a Hamiltonian flow generated by the Hamiltonian
functional $\mcn_{\lambda}(\mbA,\Phi)$, which was just defined.
(A short calculation reveals that the third equation of 
\eqref{scs} is preserved by the evolution,
and as such is really only a condition on the initial data. It will be
referred to as the constraint equation.)

\subsection{Self-dual vortices and dynamics in  the limit $\lambda \to 1$.}
\la{mainstat}
The system \eqref{scs} admits {\em soliton} solutions, 
called abelian Higgs, or Ginzburg-Landau, vortices,
which are energy minimizing critical points of the 
static energy functional $\mcn_{\lambda}(\mbA,\Phi)$.
We now discuss these solutions and their uses in understanding the dynamical
system \eqref{scs} via the adiabatic approximation.
There is a special case, $\lambda=1$, 
in which the adiabatic approximation is particularly
powerful because the space of vortices is then unusually large - large enough that 
the motion on it can provide information on the dynamical interaction
of several vortices.  We call this the self-dual, or Bogomolny,
case, and the corresponding solutions are called self-dual vortices. Now
for such a solution, $(\mbA,\Phi)$, with a
given value of the topological integer $N$, (the degree of $L$),
the field $\Phi$ will have
$N$ zeros, counted with multiplicity. 
Each of these zeros can be thought of as the centre of a vortex. Thus
the static solitons can be thought of as a nonlinear superposition of $N$ vortices
which do not interact. 
This was first fully understood
in the case that $\Sigma$ is the upper
half plane with canonical metric,
when the equations were solved exactly 
by Witten (1977) 
by reducing them to the Liouville equation. In general it is still 
possible to make a reduction
to a nonlinear elliptic equation of Kazdan-Warner type, whose solutions
can be completely parametrized although not explicitly given.
Following this,
Taubes proved an existence theorem when $\Sigma$ is the Euclidean plane
(Jaffe and Taubes 1982), and Bradlow (1988) did likewise for $\Sigma$ a compact Riemann surface,
proving the following:
\begin{theorem}[Existence of vortices on a surface,\cite{brad88}]\la{vortex}
If the area of a closed Riemann surface $|\Sigma|$ is such that
$|\Sigma|>4\pi N$ the Bogomolny bound is saturated: in fact 
the minimum value $\pi N$ of $\mcn$, where 
\begin{align}
\la{sde}
\mcn: \ch_2&\to\R\\ 
\mcn(\mbA,\Phi) &\equiv {\mcn}_{1} (\mbA,\Phi)  
= \frac{1}{2}\int_{\Sigma} \left( B^2+|{\bf D}\Phi|^2 
+\frac{1}{4}(1 -|\Phi|^2)^2
\right) \ d\mu_g,\notag
\end{align}
is achieved on a set 
$\cs_N\subset\ch_2$ of pairs $(\mbA,\Phi)$ which solve the 
{\bf Bogomolny}, or {\bf self-dual vortex}, equations:
$$
\dbar_\mbA\Phi
=0,\qquad
B-\frac{1}{2}(1-|\Phi|^2)=0.
$$
These minimizers will be referred to as the self-dual vortices, 
or just vortices.
The quotient of $\cs_N$ by the gauge group $\cg$
can be identified
with $\Sym^N(\Sigma)$, the symmetric $N$-fold product of $\Sigma$, via the
mapping which takes $\Phi$ to the set of its zeros.
\end{theorem}

\begin{remark}[Interaction and stability of vortices]
{\em The physical interpretation of theorem 
\ref{vortex} is that for $\lambda=1$ the vortices do 
not interact; see \cite{jt82} for a discussion of this, 
and some related conjectures,
and \cite{sg} for some stability theorems.}
\end{remark}
\begin{remark}[Bogomolny structure and Bogomolny operator]
\la{bogom}
{\em 
The structural feature of $\mcn$ which makes theorem 
\ref{vortex} possible was 
identified by Bogomolny in \cite{bogo76}. In this instance it amounts to the
fact that if we introduce the Bogomolny operator ${\cal B}$ to be the nonlinear
operator which maps $(\mbA,\Phi)\mapsto \left(B-\frac{1}{2}(1-|\Phi|^2),\dbar_\mbA\Phi\right)$ then 
$$
\mcn=\frac{1}{2}\int|{\cal B}(\mbA,\Phi)|^2d\mu_g
+\pi N
$$ 
(see
\S\ref{idsd} for more information in this regard). 
Also see
\cite{brad88} for higher dimensional versions of this
decomposition, and \cite{hh} for generalizations to 
solutions with non-vanishing electric field.
}
\end{remark}
\begin{remark}[Geometry of moduli space]
\la{geom}
{\em
Quotient spaces of the type arising in 
theorem \ref{vortex} are usually known as moduli spaces: in this
case we define the moduli space ${\cal M}_N$ to be
the space of gauge equivalence classes of 
self-dual vortices, so that ${\cal M}_N\equiv \Sym^N(\Sigma)$.
We call the space $\cs_N$ the vortex space and $\pr:\cs_N\to\cm_N$
the natural projection which takes $({\bf A},\Phi)$ 
to its gauge equivalence class
$[({\bf A},\Phi)]$.
The space ${\cal M}_N$ inherits both a metric
(induced from the $L^2$ metric) and a symplectic structure and
is a Kaehler manifold (see \cite{braddask91}). Explicitly, we can identify
the tangent space to $\cm_N$ with solutions $(\dot \mbA,\dot\Phi)$ 
of the linearized Bogomolny 
equations which also satisfy the condition \eqref{gauge1}.
The complex structure
and symplectic structure on $\cm_N$ are then given by restricting the formulas
given in the previous section to such $(\dot \mbA,\dot\Phi)$, and consequently
we will use the same notation, $\J$ and $\Omega$, for these objects. The
existence of this complex structure  on $ {\cm_N} $ can be seen very
clearly in the formulas in \S\ref{idsd}, in which complex
notation is used to combine the linearized Bogomolny equations with
\eqref{gauge1} into a manifestly 
complex linear operator $\mcd_\psi$,
for  $\psi=(\mbA,\Phi)\in\cs_N$.
This can all be summarized by saying that  we have an identification
\beq
\la{idm}
T_{[\psi]}{\cal M}_N
\approx\Ker\mcd_\psi
\equiv\{(\dot \mbA,\dot\Phi):DB_{\psi}[\dot \mbA,\dot\Phi]=0,\;\hbox{ and
\eqref{gauge1} holds}\}.
\eeq
}
\end{remark}
\subsection{Statement of the adiabatic limit  theorem}
\la{adlim}
In order to define the adiabatic limit system,
we now define a Hamiltonian function
$
{\cal M}_N \to\R $ 
by restricting the energy $\mcn_{\lambda}$ to the space of vortices, 
and observing that
by gauge invariance this actually gives a smooth function on the quotient
space ${\cal M}_N $. The corresponding Hamiltonian flow
determines the slow motion of vortices for 
$\lambda$ close to $1$: 
\begin{quote}
\la{cssmain}
{\em 
For $\epsilon=|\lambda-1|$ sufficiently small, the system \eqref{scs} 
can be approximated,
for times of order $\frac{1}{\epsilon}$, by
the Hamiltonian flow 
on the phase space
${\cal M}_N=\Sym^N(\Sigma)$ associated to the 
Hamiltonian function $\mcn_\lambda|_{{\cal M}_N}$ via
the symplectic form $\Omega$.
}
\end{quote}
We now move towards a precise formulation of this 
in theorem \ref{main}.
Since we are interested in the regime in which $|\lambda - 1| \ll 1$ it
is useful to introduce  a large
parameter
\begin{align}
\mu &= \frac{1}{|\lambda -1 |} \\
\intertext{and  let also, for $\lambda\neq 1$,}
\sigma  &=
  \frac{\lambda -1}{|\lambda -1|} = \pm 1
\label{lm}
\end{align}
(also defining $\sigma = 0 $ for $\lambda = 1$ where necessary).  
We  rescale time by $\tau = \frac{t}{\mu}$, and $A_0$ similarly,
leading to the following {\em rescaled equations}: 
\begin{align} 
\begin{split}
  & \frac{\partial A_1}{\partial \tau}= \mu \big(-\partial_1 B -
\langle i\Phi, D_2\Phi\rangle\big)+\frac{\partial A_0}{\partial x^1},\\ 
  & \frac{\partial A_2}{\partial \tau}= \mu \big(-\partial_2 B + 
\langle i\Phi, D_1\Phi\rangle\big)+\frac{\partial A_0}{\partial x^2},\\ 
  & i(\frac{\partial}{\partial\tau}-iA_0 ) \Phi =
\mu ( -\Delta_A\Phi -\onetwo(1-|\Phi|^2)\Phi) 
        -\frac{\sigma}{2}(1-|\Phi|^2)\Phi.
\end{split}\label{scsr}
\end{align}
It is also natural
to separate the energy 
$\mcn_{\lambda} $ into the (main) self-dual piece
$\mcn=\mcn_{1}$, and a perturbation term proportional $\lambda - 1$. 
Under the rescaling just introduced, the energy
rescales by a factor $\mu$, leading us to consider the
Hamiltonian  $H= \mu{\cal V}  + U $, where 
$\mcn\equiv\mcn_1$ is as in \eqref{sde}, and
the energy correction away from the self-dual, or Bogomolny,
regime is given by
\beq
\la{ec}
U (\Phi)   = \frac{\sigma}{8} \int_{\Sigma} (1 -
|\Phi|^2)^2\ d\mu_g .
\eeq

The rescaled equations \eqref{scsr} 
can be written as a Hamiltonian evolution for
$\psi = (\mbA,\Phi)$ in the form 
\beq\la{resceq}
\J\frac{\partial \psi}{\partial \tau} = \mu \mcn ' + U ' 
+ \J(dA_0 ,iA_0 \Phi)
\eeq
where $\J$ is the complex structure introduced at the end of \S\ref{varham},
\beq
\la{defcom}
\J(\dot A_1 dx^1+\dot A_2 dx^2, \dot\Phi) = 
(-\dot A_2 dx^1+\dot A_1 dx^2, i\dot\Phi)
\eeq
with 
$
\begin{array}{ll}
\dot \mbA = \frac{\partial \mbA}{\partial \tau}  & \dot \Phi
=\frac{\partial\Phi}{\partial \tau}.
\end{array}
$
\begin{remark}[Explicit formulation of adiabatic limit system]
\la{explicit}
{\em
We now write the equations for the adiabatic limit system in an explicit way
which will be useful later. The function $U$ is clearly gauge invariant and 
defines by restriction a smooth function $u$ on ${\cal M}_N$. Now
recall \eqref{idm}: under this identification,
the gradient of the function $u$ on ${\cal M}_N$ at $[\Psi_S]$ is identified 
with $\proj_{\Psi_S} U'$, where $\proj_{\Psi_S}$ is the orthogonal projector
onto $\Ker\mcd_{\Psi_S}$ (see lemma \ref{specproj}). 
The Hamiltonian differential equations for $u$
are then equivalent to 
\beq
\la{exh}
\J{\frac{\partial\Psi_S}{\partial\tau}}=\proj_{\Psi_S} U'.
\eeq
Given an initial value $\Psi_S(0)=\psi_0\in{\cal S}_N$, this equation
has a unique solution $\tau\mapsto\Psi_S(\tau)\in{\cal S}_N$
which satisfies the gauge condition \eqref{gauge1}.
}
\end{remark}
\begin{maintheorem}[Adiabatic limit]
\la{main}
Let $\Psi_\mu$ be the smooth solution of \eqref{resceq},
satisfying the gauge condition \eqref{gauge1}, with 
smooth initial data
$\Psi_\mu(0)$, 
such that 
\begin{enumerate}
\item[(i)]\quad
$\lim_{\mu\to+\infty}\|\Psi_\mu(0)-\psi_0\|_{{\cal H}_2}=0$, for some smooth
$\psi_0\in{\cal S}_N$, and 
\item[(ii)] \quad
$\sup_{\mu\geq 1}\|\Psi_\mu(0)\|_{{\cal H}_2}+\|\dot \Psi_\mu(0)\|_{{H}_1}\leq K<\infty.$
\end{enumerate}
Then there exists $\tau_*>0$, independent of $\mu\geq 1$, such that for 
$s<2$,
\beq\la{mains}
\lim_{\mu\to\infty}\sup_{[-\tau_*,\tau_*]}\bigl\|\Psi_\mu(\tau)
-\Psi_S(\tau)\bigr\|_{\ch_s}=0
\eeq
where $\tau\mapsto\Psi_S(\tau)\in\cs_N$ is a curve in the vortex space $\cs_N$,
also satisfying \eqref{gauge1}, which is the unique solution of
\eqref{exh} with initial data $\Psi_S(0)=\psi_0$. The
projection onto the moduli space $\cm_N$:
$$
\tau\mapsto \bigl[\Psi_S(\tau)\bigr]\in\cm_N,
$$
is the unique solution of the Hamiltonian system on  
$(\Sym^N(\Sigma),\Omega)$ 
associated to the Hamiltonian $u$ defined in remark 
\ref{explicit},
with initial value $[\psi_0]\in\cm_N$.
\end{maintheorem}
This theorem in proved in \S\ref{pmainth}, employing a strategy
which is explained in \S\ref{simple}, following discussion
of a very simple model problem. Some of the novel features 
which arise in the 
implementation of this strategy for \eqref{scsr} are highlighted  
at the begininng of \S\ref{pmainth}.

\begin{remark}[Related work]
{\em The approximation of the dynamical system
\eqref{scsr} by a dynamical system through a space of 
equilibria (in this case the self-dual vortices, which are the
equilibria for $\lambda=1$) is referred to as an adiabatic limit or
approximation. It was suggested in \cite{gobs97}, following 
earlier conjectures of the same author 
on vortex and monopole dynamics in 
second order Lorentz invariant systems discussed in 
\cite{ms2004}. Proofs of the validity of the approximation
in the case of second order dynamics were given in 
\cite{s94,s94b}; the strategy for the proof here, however, 
is different
from that adopted in those references - 
see the discussion in \S\ref{simple}. There has also been work
on corresponding problems for $\sigma$-models, see \cite{hs,rs}.
A review of the analysis of adiabatic limit 
problems is given in \cite{s07}, 
mostly directed towards infinite dimensional 
natural Lagrangian systems of the type appearing in 
classical field theory.
(Natural Lagrangian systems are those derivable from 
Lagrangians of the classical ``kinetic energy minus potential
energy'' form).} 
\end{remark}

\begin{remark}[Implications for Chern-Simons vortex dynamics]
\la{phenrem}
{\em Although it is not generally
possible to evaluate explicitly the Hamiltonian and 
symplectic form in the reduced system \eqref{exh},
it is possible to understand some basic features
of the vortex dynamics in this model, see 
\cite{ks,gobs97,rsp,ms2004}. This work has been directed mostly
to the case when the spatial
domain is $\R^2$, so our theorem \ref{main} does not imply
the validity of the approximation \eqref{exh} in this case, see below.
One general conclusion is that in the Chern-Simons model
a force acting on the vortex 
produces motion at right angles to the direction of the force (in distinction
to the behaviour in the relativistic case \cite{ms2004,s94}). Now it
is known computationally (see \cite{jt82,ms2004} and references therein),
and in some special cases analytically (\cite{s99}), that the potential energy
between two vortices depends on the distance between them, and is attractive
for $\lambda<1$ and repulsive for $\lambda>1$. 
From this it can be deduced that two vortices will circle about one another,
the direction of rotation depending upon whether $\lambda<1$ or $\lambda>1$.
See \cite[Section 7.13]{ms2004} for 
a discussion of these solutions in the $\R^2$ case. Also in the same reference
it is observed that \eqref{exh}
possesses another related type of solution: a rigidly 
rotating $p$-gon, with $p$ vortices placed at the 
vertices of a regular $p$-gon. 
Many of the arguments and calculations leading
to the conclusions about vortex dynamics can be carried out 
equally well with spatial domain the standard sphere $\Sigma=S^2$ (\cite{r}), 
even with
explicit formulae in special limiting cases (\cite{s99}),
in which case theorem \ref{main} implies rigorously the 
rotational behaviour for vortices described above. In future work
results on the existence and stability of such periodic solutions for the 
full system \eqref{scs} will be presented.

It is to be hoped that some of these
qualitative conclusions about vortex dynamics,
(which are justified for \eqref{scs} by the Main Theorem \ref{main})
would have a wider validity for Chern-Simons models of vortex dynamics,
not necessarily close to any self-dual limit.
There is some numerical
evidence for this in related situations, for example the scattering of
vortices in the relativistic abelian Higgs model is qualitatively
similar for all values of the Higgs coupling constant, even though a
rigorous analysis in which the vortices actually collide is only possible in
the self-dual limit; see \cite{s94,ms2004}). On the other hand, 
the case of first order dynamics is in some ways numerically more problematic
since it is not possible to produce any motion via choice of initial conditions
(as can be done in the second order case), 
and it is necessary to have $\lambda$ deviate 
from the self-dual value $1$, and quite substantially so in order to
get motion which is easily computationally observable.
A numerical study in \cite{ks} 
which compares the approximation \eqref{exh} with
a computer simulation of \eqref{scs} finds that,
{\em in the case of spatial domain $\Sigma=\R^2$,}
while the qualitative
behaviour of two vortices is similar to that implied by \eqref{exh}
for $|\lambda-1|$ small, there are
quantitative differences between the full dynamics and
the adiabatic limit, which become quite marked
as $\lambda$ moves away from the value $1$. 
As the authors of \cite{ks} say, 
it is unclear to what extent some of these differences
are genuine errors due to the neglect of radiation in 
the finite dimensional truncation \eqref{exh}, as compared to
being a numerical artefact; certainly some of the observed behaviour
is consistent with energy being transferred into radiative modes,
causing the vortices to spiral in towards one another in the attractive case
(\cite[Figure 6]{ks}).
In any case, there is no issue with
radiation {\em when $\Sigma$ is a compact spatial domain}, in which case
theorem \ref{main} does imply the validity of the approximation
\eqref{exh} for sufficiently small $|\lambda-1|$, and it seems
reasonable to expect that in this case
the dynamical
bevaviour predicted by our analysis (relating 
\eqref{scs} to \eqref{exh} for small $|\lambda-1|$) is at least
qualitatively relevant to the applications in the
theoretical physics literature.}

\end{remark}
\subsection{A simple model problem and discussion of methodology}
\la{simple}
We consider here a simple two-dimensional example in order to exhibit
as clearly as possible the phenomenon under study, 
and the strategy which will be employed in the proof of theorem \ref{main}.
(It is the basic strategy taken in \cite{rh57} for 
finite dimensional natural Lagrangian systems, here adapted to 
the case of infinite dimensions and to take advantage of the 
Bogomolny structure.)
For real numbers $\beta$ and $\mu {\gg}1$, we consider a linear first order Hamiltonian
system for $z(\tau)=(z^1(\tau),z^2(\tau))\in\C^2$:
\begin{theorem}
\la{thez}
For each $\mu{\gg}1$, let $\tau\mapsto Z_\mu(\tau)\in\C^2$ be the
solution of
\begin{align}
\la{ezs}
\begin{split}
\dot z^1 & = i (z^1 +\beta z^2)\\
\dot z^2 & = i (\beta z^1 +\mu z^2 ),
\end{split}\end{align}
with initial data satisfying $|(Z_\mu^1(0),Z^2_\mu(0))-(\gamma,0)|=O(\mu^{-1})$ as
$\mu\to +\infty$, for some fixed $\gamma\in\C$. Then
\beq
\la{ez}
\lim_{\mu\to+\infty}\max_{\tau\in\R}|Z_\mu(\tau)-(\gamma e^{i\tau},0)|=0.
\eeq
\end{theorem}
\begin{remark}
{\em
The system \eqref{ezs} is Hamiltonian with the standard symplectic structure on $\C^2$
and with Hamiltonian function $\mu \mcn+U$ with $\mcn(z)=\frac{1}{2}\bar z^2z^2$ and
$$
U(z)=\frac{1}{2}\bar z^1 z^1
+\beta(\bar{z}^1 z^2 + \bar{ z}^2 z^1).
$$
Thus $\mcn$ acts as a constraining potential for $\mu\to+\infty$, forcing the solution
onto the set
${\cal S} = \Complex \times \{0\} \subset\Complex^2$ where
$z^2=0$.  
Projecting the system to ${\cal S}$ gives, formally,
\begin{equation}
\label{equil}
i\dot z^1   +z^1 = 0.
\end{equation}
The theorem
asserts that \eqref{equil} indeed governs the behaviour of the
limit of appropriate sequences of  solutions to \eqref{ezs}.
}
\end{remark}
\proof
The solution
with initial data $z(0) = (z^1(0), z^2(0))$ is given by:
\begin{align}
z^1 (\tau) &= \frac{\beta}{\beta(\lambda_+ -\lambda_-)}\; 
\left[\left( (1-\lambda_-)e^{i\lambda_+\tau} 
-(1-\lambda_+)e^{i\lambda_-\tau}\right)z^1(0) +\beta\left( e^{i\lambda_+\tau}-e^{i\lambda_-\tau}\right)z^2(0)\right] \notag\\
z^2 (\tau) &= \frac{-1}{\beta(\lambda_+ -\lambda_-)}
\left[ (1-\lambda_+)(1-\lambda_-)(e^{i\lambda_+\tau}-e^{i\lambda_-\tau})z^1(0)\right]\notag\\
&\qquad\qquad\qquad\qquad
+
\frac{-\beta}{\beta(\lambda_+ -\lambda_-)}
\left[\left((1-\lambda_+)e^{i\lambda_+\tau}
- (1-\lambda_-)e^{i\lambda_-\tau}\right)z^2(0)\right].\notag
\end{align} 
Here the $\lambda_{\pm}$ are the characteristic values of the system:
$$
\lambda_{\pm} = \frac{1}{2} \left(1 + 
\mu\right)
\left[1 \pm\left(1- \frac{4(\mu -\beta^2)}{(1+\mu)^2}\right)^{\frac{1}{2}}\right],
$$
which satisfy, by the binomial expansion,
$$
|\lambda_+\,- \,\mu| = O(1),\quad
|\lambda_-\, -\,1|=O(\mu^{-1}).
$$
as $\mu\to\infty$. From this, and the fact that $\lambda_{\pm}\in\R$ for large
$\mu$ so that $|e^{i\lambda_{\pm}\tau}|=1$, the behaviour in \eqref{ez} follows
for the solutions $Z_\mu(\tau)$ with initial data as described.
\qed
\begin{remark}
\la{rt}
{\em
In this example the exact solutions indicate that while $Z^2_\mu\to 0$, the
time derivatives $\dot Z^2_\mu$ are bounded, but cannot generally be expected to have limit
zero.
}
\end{remark}

In the absence of explicit formulae for $Z_\mu(\tau)$, 
it is still possible to prove results like theorem \ref{thez}, either
\begin{enumerate}
\item[(i)]
by explicit perturbative construction of solutions to the full system, using 
solutions of the restricted system as a starting point, or
\item[(ii)]
by obtaining uniform bounds for the $Z_\mu(\tau)$ which allow the extraction of convergent
subsequences, and then identifying the unique limit of all 
such subsequences
as the corresponding solution of the restricted system with 
Hamiltonian $U\bigr|_{\cal S}$.
\end{enumerate}
In the present article we will adopt the second strategy in our proof
of theorem \ref{main}
(although it would be possible
to use the first strategy, as in \cite{s94}). 
To make the structure of the proof transparent, 
it is useful to consider in some detail how to
execute the second strategy to prove a variant of theorem \ref{thez}:
\begin{theorem}[Weaker version of theorem \ref{thez}]
\la{thez2}
In the situation of \ref{thez}
\beq
\la{ez2}
\lim_{\mu\to+\infty}\max_{a<\tau<b}|Z_\mu(\tau)
-(\gamma e^{i\tau},0)|=0,
\eeq
for every bounded interval $[a,b]\subset\R$.
\end{theorem}
\begin{remark}
{\em Although weaker than theorem \ref{thez}, the proof of 
theorem \ref{thez2} that we
give generalizes to the infinite dimensional problem \eqref{scs}, \eqref{scsr}, in which the explicit solutions corresponding to
those used in the proof of theorem \ref{thez} are of course 
not available.
}
\end{remark}
\proof
\begin{itemize}
\item
Differentiation of the equations \eqref{ezs} in time gives the identical system
$\zeta=\dot z$. Use the energy identity:
$$
\mu\mcn(\zeta(\tau))+U(\zeta(\tau))=\mu\mcn(\zeta(0))+U(\zeta(0)),
$$
together with the identical estimate for $z(\tau)$, to deduce (using Cauchy-Schwarz) that
the solutions $Z_\mu$ of theorem \ref{thez} satisfy
$|Z_\mu(\tau)|+|\dot Z_\mu(\tau)|\le C$, with $C$
independent of $\mu{\gg}1$.
\item
By the previous item, deduce that the family of functions $\tau\mapsto Z_\mu(\tau)$
is uniformly (in $\mu{\gg}1$) bounded and equicontinuous, and so the Arzela-Ascoli
theorem implies {\it subsequential} convergence $Z_{\mu_j}\to Z$ 
in $C(I)$ for any bounded interval $I\subset\R$.
\item
The energy estimate implies
that, for large $\mu$ there exists $C>0$, independent of $\mu$, 
such that $\mu\bar Z^2Z^2\leq C$. It follows that $Z^2_\mu\to 0$ along any
convergent subsequence. Now consider the integrated form of the first equation of 
\eqref{ezs} (i.e. project the system onto ${\cal S} = \Complex \times \{0\} \subset\Complex^2$ where $z^2=0$). Taking the limit $\mu_j\to\infty$, it follows that the limit 
 $Z=(Z^1,Z^2)$ of
any convergent subsequence satisfies 
$Z^1(\tau)=i\int_0^\tau Z^1(\tau')d\tau'$ and $Z^1(0)=\gamma$. This integral
equation has unique solution $Z^1(\tau)=\gamma e^{i\tau}$, and hence the $C_{loc}$ limit
of any convergent subsequence is $(\gamma e^{i\tau},0)$. It follows that
$Z_\mu$ converges to this limit in $C_{loc}$ without restriction to subsequences.
This proves theorem \ref{thez2}. 
(In view of remark \ref{rt} we should not expect this
convergence to be in $C^1_{loc}$.)\qed
\end{itemize}

The general situation to which theorem 
\eqref{thez2}, and its proof, potentially generalize is the
following: on a phase space ${\cal H}$ we consider the integral 
curves $Z_\mu(\tau)$  for a Hamiltonian
$\mu\mcn+U$ for large $\mu$ (``the full system''). Under the 
assumption that 
${\cal S}=\{z\in{\cal H}:\min\mcn=\mcn(z)\}$ is a symplectic 
submanifold of 
${\cal H}$, we can consider the ``restricted system'' 
on ${\cal S}$ determined
by the Hamiltonian $U\bigr|_{\cal S}$, 
and try to prove that this Hamiltonian system can be used to
describe the limiting behaviour of $Z_\mu(\tau)$ as $\mu\to +\infty$. 
An infinite dimensional example of this situation is provided
by the Chern-Simons-Schr\"odinger system \eqref{scsr}:
in the next section we will provide a proof of theorem \ref{main} 
employing the same strategy to that used in the proof of theorem
\ref{thez2} just given.

\section{\large{Uniform bounds and proof of the main theorem}}
\label{pmainth}

In this section we prove our main result, theorem \ref{main}, 
along the lines suggested
by the discussion of the simple model problem in the 
last section.
The crucial stage is
the proof of the main estimate, theorem \ref{mest}, 
which asserts the existence of 
a time interval, {\it independent of $\mu$}, on which the solution 
$\psi=(\mbA,\Phi)$
is uniformly bounded
in $\ch_2$, and its time derivative is uniformly bounded in $H^1$
as $\mu\to+\infty$. Given this bound,
theorem \ref{main} can be deduced using a variant of the
Lions-Aubin lemma, and a careful analysis of the
$\mu\to+\infty$ limit of \eqref{scsr}. 
Before obtaining the uniform bound, we collect some 
identities used in the proof.  Some more specialized identities 
related to 
the self-dual structure are collected separately in 
\S\ref{idsd}, and
referred to as needed. Specifically, we draw the reader's attention
to the following two uses made of these more specialized identities:
\begin{enumerate}
\item[(i)] Differentiation in time gives rise to an equation \eqref{dresceq}
for $\zeta=\dot\psi$ in which the dominant
term  (as $\mu\to+\infty$) involves 
${\overline L}_\psi$, the Hessian of $\mcn$ defined in
\eqref{modhess}. It is shown in   
\S\ref{idsd} that this operator takes the special form  
\beq\la{disp}
{\overline L}_\psi  = \mcdpsi^{\ast}\mcdpsi+O(|{\cal B}|),
\eeq
with $\mcdpsi$ complex linear (see \eqref{jlin}), and
$\mcb$ as in remark \ref{bogom}.
Observing that the $L^2$ norm is 
exactly preserved for equations of the form
$\J\dot\zeta=\mcdpsi^{\ast}\mcdpsi\zeta$, it is easy to believe
that the stated structure of ${\overline L}_\psi$ is useful
in the derivation of $\mu$-independent bounds for \eqref{dresceq},
(for initial data as in the theorem); this  indeed turns out to be the case - see the proof of theorem \ref{mest}.
\item[(ii)] After obtaining a convergent subsequence of solutions of \eqref{resceq} it is necessary to
take the limit of the equation  itself along the subsequence
$\mu=\mu_j\to+\infty$. For this purpose it is very convenient to be able 
to eradicate the term $\mu\mcn'$ on the right hand side, since this is
clearly hard to control for large $\mu$: this can be done by applying 
a projection operator $\proj_\mu$ whose existence close to the set
of self-dual vortices is assured by the Bogomolny structure: see
lemmas \ref{w} and \ref{specproj}. (In geometrical terms there
is a foliation of the phase space $\ch_2$, and the range of $\proj_\mu$
is the tangent space to the leaves of this foliation,
after dividing out by the action of the gauge group
using \eqref{gauge1}.)
\end{enumerate}

Although our final conclusions are in terms
of the standard Sobolev norms based on the fixed connection $\nabla$, 
it will be convenient to obtain bounds for the corresponding Sobolev norms
defined at each fixed time with respect to the connection 
${\bf D}=\nabla-i\mbA$, see \eqref{dhonea}. 
These can be related to the standard
norms by \eqref{end1}-\eqref{end3}.
\subsection{The evolution equations and associated identities}\label{eveqid}
In addition to 
the rescaled equation \eqref{resceq}
for $\psi = (\mbA,\Phi)$:
\beq
\J\frac{\partial \psi}{\partial \tau} = \mu \mcn ' + U ' + \J(dA_0 ,
iA_0 \Phi),
\notag\eeq
we will use the differentiated equation for 
$\zeta = \dot\psi \equiv \frac{\partial \psi}{\partial\tau}$. To write this down
we need the  linearization of the operator $\mcn ' (\psi) $, i.e.
the second order linear differential operator $L_\psi$ obtained by
differentiation of the map $\psi\mapsto\mcn'(\psi)$:
$$
L_\psi = D\mcn ' (\psi),
$$
or equivalently, $\langle\zeta,L_\psi\zeta\rangle_{L^2}=
\frac{d^2}{ds^2}\mcn(\psi+s\zeta)|_{s=0}$. Explicitly, with
$\zeta=(\dot \mbA,\dot\Phi)$, we have
\begin{align}\la{Lr}
\langle\zeta,L_\psi\zeta\rangle_{L^2}=\int\biggl(
|\mathbf{d}\dot \mbA|^2+|D\dot\Phi|^2+|\Phi|^2|\dot \mbA|^2 
-2\langle {\mathbf D}\Phi,&i\dot \mbA\dot\Phi\rangle
-2\langle {\mathbf D}\dot\Phi,i\dot \mbA\Phi\rangle \\
&+\langle\Phi,\dot\Phi\rangle^2-\frac{1}{2}(1-|\Phi|^2)|\dot\Phi|^2\biggr)
d\mu_g.\notag
\end{align}
\begin{remark}
{\em
There is a slightly simpler version of this formula, given in 
\eqref{modhess} below, when $\zeta$ is restricted
by the gauge condition \eqref{gauge1}. Furthermore 
in \S\ref{idsd} it is shown that the self-dual structure
provides a useful way of rewriting this formula 
as in \eqref{disp}, in terms of the complex 
structure defined in \eqref{defcom}, and using the complex one-form
 $\dot\alpha dz$, where $\dot\alpha = \frac{\dot
  A_1 -i\dot A_2}{2}$, in place of the real one-form 
$\dot A_1dx^1+\dot A_2 dx^2$, see \eqref{hesssd}.
Since this is used only at one point in the proof - in lemma
\ref{estzl2} - this formulation is presented separately 
in \S\ref{idsd}, and
referred to only as needed.
}
\end{remark} 
The linearization of $U ' $ is the linear operator
$K_\psi=D U'(\psi)$, given by
\beq\la{Kr}
K_\psi = (\dot \mbA, \dot \Phi) \mapsto \left (0, \frac{\sigma}{2}
(1-|\Phi|^2)\dot\Phi +\sigma \langle\Phi, \dot\Phi\rangle\Phi\right),
\eeq
with $\sigma$ defined in \eqref{lm}.
Given these definitions, the chain rule implies that, 
if $\psi$ is a smooth solution of \eqref{resceq},
then $\zeta(\tau)=\dot\psi(\tau)$ solves
\beq\la{dresceq}
\J\frac{\partial\zeta}{\partial\tau } 
= \mu L_\psi \zeta +K_\psi\zeta
+\J\frac{\partial}{\partial\tau }(dA_0, iA_0 \Phi).
\eeq



We also need identities for the evolution of the Bogomolny operator $\mcb$
defined in remark \ref{bogom} and discussed in more detail in \S\ref{idsd}.
The first component is preserved
\beq\la{lincon}
\frac{\partial}{\partial \tau }\bigl((B-\onetwo(1 - |\Phi|^2)\bigr) = 
e^{-2\rho} (\partial_1 \dot{A}_2 - \partial_2 \dot{A}_1) +
\langle\Phi,\dot\Phi\rangle = 0,
\eeq
as a consequence of \eqref{scsr}. We will require that the initial data
are such that $B-\onetwo(1 - |\Phi|^2) = 0$ initially, and hence for all
times.
The second component of the Bogomolny operator $\mcb$ will be denoted
\beq\la{defeta}
\eta = \barpbfa\Phi=\frac{1}{2}(D_1+iD_2)\Phi,
\eeq
(see \S\ref{idsd}), and we have the following identity:
\begin{equation}\label{eqeta}
i(\partial_\tau - iA_0)\eta = \mu (-4\barpbfa (\er 
\partial_\mbA \eta)+|\Phi|^2\eta)- 
                            \frac{\sigma}{2}\barpbfa\left( (1-|\Phi|^2)\Phi\right).
\end{equation}
(To verify this identity: substitute
$\Delta_{\mbA}\Phi=4e^{-2\rho} \pbfa\barpbfa\Phi-B\Phi$ into the third line of
\eqref{scsr}
and then apply $\bar\partial_\mbA$ to the resulting equation and use the identity
$(E_1+iE_2)\Phi=-2\mu|\Phi|^2\barpbfa\Phi$ which follows from the first
two lines of \eqref{scsr}.)

Of course, the energy
\beq
\mce(\tau)=\mu\mcn(\psi(\tau))+U(\psi(\tau))=\mce_0>0
\la{eqen}\eeq
is independent of time $\tau$ for regular solutions,
as is the $L^2$ norm
\beq\la{eqp}
\|\Phi(\tau)\|_{L^2}=L>0.
\eeq
\subsection{Choice of gauge condition and related estimates}\la{gl}
The divergence of $E$ can be calculated to be:
\begin{align*}
\dv E 
  & = \er (\partial_1 E_1 + \partial_2 E_2) \\
  & = \mu \Bigl( (-\Delta B - \er \partial_1 \langle i\Phi, D_2 \Phi\rangle 
+\er \partial_2 \langle i\Phi, D_1\Phi\rangle \Bigr)\\
  & = \mu (4\er|\eta|^2 ) + \langle i\Phi, (\pt - iA_0 )\Phi\rangle 
- \sigma B|\Phi|^2.
\end{align*}
In the last line we have used
$ B = \onetwo (1- |\Phi|^2) $, so that
$\Delta B=-\langle\Phi,\Delta_\mbA\Phi\rangle
-e^{-2\rho}(|D_1\Phi|^2+|D_2\Phi|^2)$,
the equation for $\Phi$ and the definition of $\eta$ in \eqref{defeta}.
Under the gauge condition \eqref{gauge1}
we get the following equation for $A_0$:
\beq\la{eqa0}
(-\Delta +|\Phi|^2) A_0 = 4\mu \er|\eta|^2 - \sigma B|\Phi|^2.
\eeq
\begin{lemma}[Estimates for $A_0$]
\la{esta0}
Assume 
$\tau\mapsto\psi(\tau)=(\mbA(\tau),\Phi(\tau))$, 
is a smooth solution,
of \eqref{resceq} which 
satisfies the gauge condition \eqref{gauge1},
\eqref{eqen}  and \eqref{eqp}. Then
for all $r<\infty$, there
exists $c_0({\cal E}_0,L,r)>0$ such that, 
\begin{align}
\|A_0 (\tau)\|_{L^r} &\leq \ c_0(\mce_0,L,r)\\
\intertext{and there exists $c_0(\mce_0,L)>0$ such that}
\|A_0 (\tau)\|_{H^2} &\leq \ c_0(\mce_0,L)
(1+\mu\|\bar\partial_\mbA\Phi(\tau)\|_{L^\infty}).
\end{align}
\end{lemma}
\begin{remark}
{\em
This shows that in the original system (before rescaling)
the time component of the potential $A_0$ is $O(|\lambda-1|)$ 
in the gauge defined by \eqref{gauge1}.
}
\end{remark}
\proof
The crucial point here is the $\mu$ independence of the bounds.
The second inequality follows from standard elliptic theory once
the first is established.
By \eqref{eqa0} it is possible to write $A_0=A_0^+ +\hat A_0$
where
$(-\Delta +|\Phi|^2) A_0^+ = 4\mu \er|\eta|^2$,
so that $A_0^+\geq 0$ by the maximum principle, and 
$(-\Delta +|\Phi|^2) \hat A_0 = -\sigma B|\Phi|^2$.
The bounds stated in the lemma will follow by the triangle
inequality once they are proved for $A_0^+$, since they
are immediate for $\hat A_0$. Now integrating the
equation for $A_0^+$ implies that
$\||\Phi|^2 A_0^+\|_{L^1}=\int_\Sigma |\Phi|^2
A_0^+d\mu_g\leq C(\mce_0,L)$
since $A_0^+\geq 0$; this bound is independent of $\mu\gg 1$ on
account of \eqref{eqen}. The standard elliptic theory for
$-\Delta u=f\in L^1$ now gives the $L^r$ estimates for
$A_0^+$ and hence the lemma.
\qed
\begin{lemma}[Estimates for $\dot \mbA$]
\la{esta}
Let $\zeta=(\dot \mbA,\dot\Phi)$
satisfy the gauge condition \eqref{gauge1},
as well as the linearized constraint equation
\eqref{lincon}.
Then there exists a constant $c_1>0$ such that
$\|\dot \mbA\|_{H^1}\le c_1\|\Phi\dot\Phi\|_{L^2}$, and 
more generally, for any
$1<p<\infty$, there exists a constant $c_1(p)>0$ such that 
$\|\dot \mbA\|_{W^{1,p}}\le c_1\|\Phi\dot\Phi\|_{L^p}$. In particular
these estimates hold for a smooth solution, 
$\tau\mapsto\psi(\tau)=(\mbA(\tau),\Phi(\tau))$, 
of \eqref{resceq} which 
satisfies the gauge condition \eqref{gauge1}.
\end{lemma}
\proof
These are the standard estimates for the Hodge system, proved by using the
Hodge decomposition to reduce to the Calderon-Zygmund estimate for the
Laplacian.\qed

On the subspace of $\zeta=(\dot \mbA,\dot\Phi)$ satisfying
the gauge condition \eqref{gauge1}, 
the operator $L_\psi$ has a simpler form: 
$L_\psi\zeta=\overline L_\psi\zeta$,
where $\overline L_\psi$ is the operator defined by 
\begin{align}\label{modhess}
\langle\zeta,\overline L_\psi\zeta\rangle_{L^2}=\int\biggl(
|\mathbf{d}\dot \mbA|^2+|\dv\dot \mbA|^2
+|{\bf D}&\dot\Phi|^2+|\Phi|^2(|\dot \mbA|^2 +|\dot\Phi|^2)\\
&-4\langle{\bf D}\Phi,i\dot \mbA\dot\Phi\rangle
-\frac{1}{2}(1-|\Phi|^2)|\dot\Phi|^2\biggr)
d\mu_g.\notag
\end{align}

\begin{lemma}[The Hessian]
\la{saop}
Let $\psi=(\mbA,\Phi)$
be smooth. Then the second order differential
operator $\overline L_\psi$ is a self-adjoint operator
with domain $H^2$, and there exist numbers $c_2,c_3$ such that
$$
\langle\zeta,\overline L_\psi\zeta\rangle_{L^2}\geq
c_2\|\zeta\|_{\honea}^2-c_3\|\zeta\|_{L^2}^2.
$$ 
The numbers $c_2,c_3$ depend only on the numbers $L$ and 
$\mce_0$, defined as in \eqref{eqen},\eqref{eqp}.
\end{lemma}
\proof
First of all, observe that
$$
\int\biggl(
|\mathbf{d}\dot \mbA|^2+|\dv\dot \mbA|^2
+|{\bf D}\dot\Phi|^2+|\Phi|^2(|\dot \mbA|^2 +|\dot\Phi|^2)
\biggr)d\mu_g\geq
c(\mce_0,L)\|(\dot \mbA,\dot\Phi)\|_{\honea}^2.$$
This can be proved by a straightforward contradiction
argument that is very similar to the proof of lemma
\ref{bdq} given below, so the details will be omitted.
Next, to deduce the stated result, just bound
the final two terms in \eqref{modhess}
using the Holder inequality with $1=\frac{1}{2}+
\frac{1}{4}+\frac{1}{4}$, the interpolation inequality
in lemma \ref{covgn} and Cauchy-Schwarz.
\qed
\begin{corollary}
Assume given a smooth solution, 
$\tau\mapsto\psi(\tau)=(\mbA(\tau),\Phi(\tau))$, 
of \eqref{resceq} which 
satisfies the gauge condition \eqref{gauge1},
\eqref{eqen} and \eqref{eqp}. Then the quantity
\beq\la{defen1}
\mce_1 (\tau) = \frac{1}{2} \langle\zeta(\tau), 
(L_\psi + \mu^{-1}K_\psi)\zeta(\tau)\rangle_\ltwo,
\eeq
where $\psi=\psi(\tau)$, satisfies for $\mu\geq 1$ 
$$
\mce_1 \ \geq \ c_4 \|\zeta\|^{2}_{\honea} - c_5 \|\zeta\|^{2}_\ltwo
$$
with $c_4,c_5$ depending only on $\mce_0,L$.
\end{corollary}
\subsection{The main estimate}\label{pme}
We say that a smooth solution, 
$\tau\mapsto\psi(\tau)=(A(\tau),\Phi(\tau))$, of 
\eqref{resceq} satisfies conditions (AE) and (AI), 
if the following conditions hold:
\begin{itemize}
\item[(AE)]
\begin{quote}\noindent
There exists positive numbers $\mce_0,L$ such that 
$\|\Phi(\tau)\|_{L^2}=L$ and $\mce(\tau)=\mce_0$, for all
times $\tau\in\Real$, where $\mce(\tau)$ is the 
energy \eqref{eqen}. (Recall that both these quantities are 
independent of $\tau$.)
\end{quote}
\item[(AI)]
\begin{quote}\noindent
The initial data are such that 
$\|\psi(0)\|_{\ch_2}+\|\dot\psi(0)\|_{H^1}\leq K<\infty$.
(Recall
the definition of the norms in \eqref{h2}). 
\end{quote}
\end{itemize}

\begin{theorem}
\la{mest}
For $\mu\geq 1$ let $\tau\mapsto\psi(\tau)$ be a smooth 
solution of \eqref{resceq} satisfying conditions (AE) 
and (AI), for some fixed numbers $K,L,\mce_0$. 
There exist numbers $\tau_*>0$ and $M_*>0$, independent of $\mu$,
such that
\begin{equation}
\label{h3}
\max_{|\tau|\leq\tau_*}
\Bigl|\bigl(\psi(\tau), \frac{\partial}{\partial\tau} \psi(\tau)
\bigr)\Bigr|_{\ch_2 \times H^1} \ \leq
\ M_*.
\end{equation}
\end{theorem}

\noindent{\it Beginning of proof of theorem \ref{mest}.}\ \ By
time reversal invariance it is sufficient to prove the
bound for $0\leq\tau\leq\tau_*$, for some $\tau_*>0$
independent of $\mu$.
Let
$$
\zeta(\tau)=\frac{\partial}{\partial\tau} \psi(\tau)=\dot\psi(\tau).
$$
For any $M> \|\zeta(0)\|_{L^2}$ there exists a 
time $T(M,\mu) >0$ such that 
\beq
\sup_{0\leq \tau \leq T(M,\mu)} \|\zeta(\tau)\|_{L^2} \leq \ M .
\la{AM}\eeq
We will prove that there exist positive numbers 
$M_*,\tau_*$, independent of 
$\mu$, such that $T(M_*,\mu)\geq\tau_*$, and hence 
$\sup_{0\leq \tau \leq\tau_*} \|\zeta(\tau)\|_{L^2} 
\leq \ M_* $. 
The proof
proceeds by obtaining a series of $\mu$-independent bounds,
predicated upon \eqref{AM},
which imply boundedness
of $\bigl(\psi(\tau),\dot\psi(\tau)\bigr)$ 
in the Hilbert space $\ch_2$ defined in \eqref{h2}
for $0\leq\tau\leq\tau_*$. 
These bounds are
now stated in a sequence of lemmas, all of which refer to a smooth solution 
of \eqref{resceq},\eqref{gauge1} which verifies (AE), (AI) and \eqref{AM}
for all  $\tau$ under consideration.
\begin{lemma}[Estimate for $\Phi$ in $H^2$]
\la{pl2}
There exists $ C_1=C_1(\mce_0,L)>0$, independent of $\mu$, such that 
$$
\|\Phi(\tau)\|_{H^2_A}
\leq C_1(1+ \|\zeta(\tau)\|_\ltwo)\leq C_1(1+M).
$$
\end{lemma}
\proof
Using the third equation of \eqref{scsr} for $\Phi$, we bound
$$
\|\Delta_{\mbA}\Phi\|_{L^2}\leq
\|\dot\Phi\|_{L^2}+\|A_0\Phi\|_{L^2}+\frac{1}{2}\|\Phi(1-|\Phi|^2)\|_{L^2}.
$$
Now, by lemma \ref{lemgarding}, we can bound
$
\|\nabla_\mbA\nabla_\mbA\Phi\|_{L^2 } \leq 
\|\Delta_\mbA\Phi\|_{L^2}
+ c(\mce_0)\|\nabla_\mbA\Phi\|_{L^4},
$
and hence, by lemma \ref{covgn} and Cauchy-Schwarz:
$
\|\nabla_\mbA\nabla_\mbA\Phi\|_{L^2 } \leq 
2\|\Delta_\mbA\Phi\|_{L^2}
+ c(\mce_0,L).
$
Therefore, using also lemma \ref{esta0},
we deduce the bound
$
\|\Phi(t)\|_{H^2}\leq c
(1+\ \|\zeta(\tau)\|_{\ltwo})\leq c(1+M),
$
for some $c= c(\mce_0,L)>0$, and the result follows.
\qed
\begin{corollary} 
\la{pinf}
$\exists C_2=C_2(\mce_0,L)>0$ such that,
$\|\Phi(\tau)\|_\linf
\leq C_2\bigl(1+\sqrt{\ln(1+M)}\bigr).$
\end{corollary}
\proof This follows from lemma \ref{covbg} and the previous lemma.\qed
\begin{lemma}[Energy estimate for 
$\zeta=\dot\psi$]
There is a constant  $C_3(\mce_0,L)>0$ 
such that,
\begin{equation}
\label{zeta}
\left|\frac{d\mce_1}{d \tau}\right| 
\leq C_3(1+\|\Phi\|_{L^\infty}^2)\|\zeta\|_{\honea}^2+
C_3\|\zeta\|_{L^2}^6+C_3\|\zeta\|_{L^2}^4.
\end{equation}
where $\mce_1$ is the quantity defined in \eqref{defen1}.
\end{lemma}
\proof
Compute $\frac{d}{dt}\mce_1$, substitute from 
\eqref{dresceq}, and use the observation that 
\beq\la{deq}
\langle \J\dot \zeta , 
(d\dot{A}_0, i\Phi\dot{A}_0 )\rangle_{L^2} = 0, 
\eeq
by the constraint equation  $B=\frac{1}{2}(1-|\Phi|^2)$
in \eqref{scs}, to obtain
$$
\frac{d\mce_1}{d \tau} = \langle i\dot\Phi, iA_0 \dot \Phi\rangle_{L^2} 
+ \frac{1}{2}
\langle \zeta, [\frac{\partial}{\partial\tau} , L_\psi+\mu^{-1}K_\psi] \zeta\rangle_{L^2}.
$$
To handle the second term, we make use of the following 
bounds (written
schematically, i.e. suppressing indices and inner products 
which play no role):
\begin{align*}
\|\Phi\zeta^3\|_{L^1}&\leq \|\Phi\|_{L^\infty}\|\zeta\|_{L^2}\|\zeta\|_{L^4}^2
\leq c\|\Phi\|_{L^\infty}\|\zeta\|_{L^2}^2\|\zeta\|_{\honea}\\
\|\dot\Phi\dot \mbA\nabla_\mbA\dot\Phi\|_{L^1}
  &\leq \|\nabla_\mbA\dot\Phi\|_{L^2}\|\dot \mbA\|_{L^4}\|\dot\Phi\|_{L^4}
\leq c\|\Phi\|_{L^\infty}
\|\zeta\|_{\honea}^{3/2}\|\zeta\|_{L^2}^{3/2}\\
\|\dot\Phi^2\nabla\dot \mbA\|_{L^1}
  &\leq\|\zeta\|_{L^4}^2\|\nabla\dot \mbA\|_{L^2}
\leq c\|\Phi\|_{L^\infty}\|\zeta\|_{L^2}^2\|\zeta\|_{\honea}.
\end{align*}
All of these bounds follow directly from Holder's 
inequality, the interpolation inequality in 
lemma \ref{lemcovgn}, lemma \ref{esta} and the bound
$$\|\dot \mbA\|_{L^4}+\|\dot \mbA\|_{H^1}\leq 
c\|\Phi\|_{L^\infty}\|\dot\Phi\|_{L^2}.$$
It then follows, by inspection of the formulae for 
$L_\psi, K_\psi$ in \eqref{Lr} and \eqref{Kr}, 
that the second term in 
$\frac{d\mce_1}{d\tau}$ can be bounded by a sum of terms 
of this type, and hence:
$$
\Bigl|\bigl\langle \zeta,\ \  [\frac{\partial}{\partial\tau}\  ,\  L_\psi+\mu^{-1}K_\psi] \zeta\bigr\rangle_{L^2}\Bigr|
\leq c(1+\|\Phi\|_{L^\infty}^2)\|\zeta\|_{\honea}^2+
c\|\zeta\|_{L^2}^6+c\|\zeta\|_{L^2}^4.
$$
Also, we can bound 
$$
| \langle i\dot\Phi, iA_0 \dot \Phi\rangle_{L^2}|\leq c
\|A_0\|_{L^r}\|\dot\Phi\|_{L^{2r'}}^2\leq c
\|A_0\|_{L^r}\|\dot\Phi\|_{\honea}^2
$$
where $r>1$ and $1/r+1/r'=1$.
Combining these with lemma \ref{esta0}, we obtain \eqref{zeta},
completing the proof of the lemma.
\qed
\begin{corollary}\la{vesth1}
There is a constant  $C_4=C_4(\mce_0,K,L,M)>0$ 
such that, $\|\zeta(\tau)\|_{\honea}\leq C_4(1+\tau)$, for all times 
$\tau\in[0,T(M,\mu)]$.
\end{corollary}
\begin{lemma}[Estimate for $\eta = \barpbfa\Phi$]
There exists $C_5=C_5(\mce_0)>0$ such that, 
at each time $\tau$,
\begin{equation}
\label{eta}
{\mu} \|\eta\|_{H^2_A} \leq
C\bigl(\|\dot\Phi\|_{\honea}+\|\dot \mbA\|_{L^2}^2+\|\Phi\|_{L^\infty}^2\bigr).
\end{equation}
\end{lemma}
\proof
From  
the equation \eqref{eqeta} for $\eta$, and using the interpolation inequality 
in lemma \ref{covgn}, the elliptic term 
$$
\mcl_{(\mbA,\Phi)} \eta \equiv (-4\barpbfa (\er \pbfa \eta)+|\Phi|^2\eta)
$$
satisfies, for some $c=c(\mce_0)>0$,
\begin{equation}
\label{ell}
{\mu}\|\mcl_{(\mbA,\Phi)} \eta \|_{L^2} \leq 
\|\dot\Phi\|_{H^1_A}+\|\Phi\|_\linf\|\dot \mbA\|_{L^2}
+c\|A_0\|_{L^4}(1+\|\eta\|_{H^1}^{1/2})+c\|\Phi\|_{L^\infty}^2.
\end{equation}

We next see that \eqref{eta} follows from the usual
elliptic regularity estimate.  Firstly, 
observe that associated to the operator
$\mclaphi$ is the quadratic form 
$$
Q_{(\mbA,\Phi)}(\eta) \ =
\ \langle\eta, \mclaphi\eta\rangle_{L^2
  (\Sigma)}
=\  \int_\Sigma \left(4|\pbfa\eta|^2 e^{-4\rho}
+|\Phi|^2|\eta|^2\er\right)\ \dmug,
$$
which is bounded below by $c\|\eta\|_{\honea}^2$ where
$c=c(\mce_0,L)>0$ by lemma \ref{bdq}. It follows that
$\|\eta\|_{\honea}\leq c\| \mclaphi\eta\|_{L^2}$, a result
which can be strengthened by the following\\
\noindent
{\it Claim:}
$
\|\nabla_\mbA\nabla_\mbA\eta\|_{L^2}\ 
\leq c\|\mclaphi \eta\|_{L^2}\ \mbox{where}\ c=c(\mce_0,L)>0.
$\\
\noindent
By the Garding inequality
$$
\|\nabla_\mbA\nabla_\mbA\eta\|_{L^2}\leq
 \|\mclaphi \eta\|_{L^2}+c(\mce_0,L)(\|\nabla_\mbA\eta\|_{L^4}
+\|\eta\|_{\honea}).
$$
Finally, using the interpolation inequality \eqref{covgn} and 
the Cauchy-Schwarz inequality, we deduce the inequality 
claimed.

\qed
\begin{corollary}
\la{iinf}
There is a constant  $C_6=C_6(\mce_0,K,L,M)>0$ 
such that, $\mu\|\barpbfa\Phi(\tau)\|_{\linf}\leq 
C_6(1+\tau)$.
\end{corollary}
\begin{lemma}[Closing the argument: estimate for $\zeta$ in $L^2$] 
\la{estzl2}
There is a constant $C_7(\mce_0,L,M)$
such that  $\zeta = \frac{\partial \psi}{\partial \tau}$  
satisfies
$$
\|\zeta (\tau)\|_{\ltwo}^{2} \leq 
 \|\zeta(0)\|_{\ltwo}^{2}e^{C_7\int_0^\tau(\|\mu\barpbfa\Phi(s)\|_\linf
+\|\Phi(s)\|^{2}_{\linf})ds}. 
$$
\end{lemma}
\proof
Compute, using \eqref{dresceq}, that 
$$
\frac{d}{d\tau}\|\zeta(\tau)\|_\ltwo^{2} \ 
= \ 2\langle \J\zeta, (\mu L_\psi +K_\psi)\zeta\rangle
$$
since (by the gauge condition)
$
\langle\zeta , (d\dot{A}_0, i\Phi\dot{A}_0 )\rangle_{L^2} = 0,
$
and 
$
\langle \zeta ,\ (0, iA_0\dot \Phi)\rangle_{L^2} = 0 
$
(using $\langle i\dot\Phi , \dot \Phi\rangle = 0$ pointwise).
By corollary \ref{c31} and the formula for $K_\psi$, there exists $C_7=C_7(\mce_0,L)>0$ such that
$$
\left| \frac{d}{d\tau}\|\zeta(\tau)\|_\ltwo^{2} \right| \leq 
C_7(\mu\|\barpa\Phi(\tau)\|_\linf
+\|\Phi(\tau)\|^{2}_{\linf})\|\zeta(\tau)\|_\ltwo^{2}
$$
and so the stated inequality follows by the Gronwall lemma.
\qed

\noindent{\it Completion of proof of theorem \ref{mest}.}
The previous lemma allows us to 
validate the claim that \eqref{AM}, and thus all the bounds 
in lemmas \ref{pl2}-\ref{estzl2},
in fact hold on a $\mu$-independent interval $[0,\tau_*]$,
thus closing the argument. Indeed,
by corollaries \ref{pinf} and \ref{iinf} we have
$\mu\|\barpa\Phi(\tau)\|_\linf
+\|\Phi(\tau)\|^{2}_{\linf}\leq C_8(1+\tau)$
for some $C_8=C_8(\mce_0,L,M)$. 
Now let $\tau_*,M_*$ be such that
$$
\|\zeta(0)\|_{L^2}^2e^{C_7C_8(\tau_*+\tau_*^2/2)}\leq M_*^2.
$$
(This is always possible for $M_*>\|\zeta(0)\|_{L^2}$ 
and $\tau_*$ small.)
Then it follows that \eqref{AM} holds with 
$T(M_*,\mu)\geq\tau_*$, and that the
bounds given in lemma \ref{pl2} through corollary \ref{iinf} hold
on the interval $[0,\tau_*]$. 
To conclude, we explain how to derive the bounds in \eqref{h3}.
For $\zeta=\dot\psi$ we have boundedness of 
$\|\zeta(\tau)\|_{\honea}$ by corollary \ref{vesth1}. Integrating in $\tau$
gives the bound for $\|\mbA\|_{\hone}$ in \eqref{h3}. Also the Kato and Sobolev
inequalities (\cite{jt82})
give a bound for $\dot\Phi$ in $L^p$ for $2\leq p<\infty$. Together
with the boundedness of $\|\Phi\|_{L^\infty}$ this implies boundedness of 
$\|\dot\mbA\|_{W^{1,p}}$ by lemma \ref{esta}. Hence, integrating in $\tau$
and applying Sobolev's inequality we deduce boundedness of $\|\mbA\|_{L^\infty}$.
Putting all this information into
\eqref{end1},\eqref{end2} we can deduce, from lemma \ref{pl2}
and corollary \ref{vesth1}, that
$\bigl(\Phi(\tau),\dot\Phi(\tau)\bigr)$ is bounded in the
($\tau$-independent) norm $H^2\times H^1$ as claimed in \eqref{h3}.
\qed
\subsection{Proof of theorem \ref{main}}\la{pmt}
There are three stages to the proof:
\begin{itemize}
\item Deduce, from the uniform bounds of theorem \ref{mest} and
the compactness lemma \ref{lal}, 
that for any sequence $\mu_j\to +\infty$, there
exists a subsequence along which the $\Psi_{\mu_j}$ converge.
\item Identify the limit of these convergent subsequences.
\item Deduce, from the uniqueness of the limit just identified, that 
the $\Psi_\mu$ do in fact converge as $\mu\to +\infty$ (without
restriction to subsequences).
\end{itemize}

The first stage of the proof depends upon the following
version of the Lions-Aubin compactness 
lemma (see \cite[lemma 10.4]{mb}),
which is proved by a modification of the standard
proof of the usual Ascoli-Arzela theorem:
\begin{lemma}
\la{lal}
Assume that $(V,h)$ is a smooth vector bundle with inner product,
over a compact Riemannian manifold $(\Sigma,g)$, which is endowed
with a smooth unitary connection $\nabla$ and corresponding Sobolev 
norms $\|\,\cdot\,\|_{H^s}$ on the space of sections defined
as in \cite{palais68}.
Assume that $l,s$ are positive numbers with $l<s$. 
Assume $f_n(\tau)$ is a sequence of smooth time-dependent sections 
of $V$ which satisfy
$$\max_{|\tau|\le {\tau}_*}\bigl(\|f_n(\tau)\|_{H^s}+
\|\dot f_n(\tau)\|_{H^l}\bigr)\le C.$$
Then there exists a subsequence $\{f_{n_j}\}_{j=1}^\infty$ which
converges to a limiting time-dependent section
$f\in C([-{\tau}_*,{\tau}_*]; H^s(V))$, in the sense
that,
$\max_{|\tau|\le {\tau}_*}
\|(f_n(\tau,\,\cdot\,)-f(\tau,\,\cdot\,))\|_{H^r}\to 0,$
for every $r<s$.
\end{lemma}
\noindent
Applying this we infer immediately the existence of a subsequence 
$\mu_j\to+\infty$ along which the solutions 
$\Psi_{\mu_j}=(\mbA^{\mu_j},\Phi^{\mu_j})$ 
converge to a limit 
$\Psi_S(\tau)$ in the sense that
\beq
\la{convs}
\lim_{\mu_j\to\infty}\sup_{[-\tau_*,\tau_*]}\bigl\|\Psi_{\mu_j}(\tau)
-\Psi_S(\tau)\bigr\|_{\ch_{r}}=0,
\eeq
for $r<2$. It follows from corollary \eqref{iinf}, that
$$\lim_{\mu\to+\infty}\sup_{[-\tau_*,\tau_*]}
\|\bar\partial_{\mbA^{\mu}}\Phi^{\mu}\|_{L^\infty}=0,
$$
and since the other Bogomolny equation 
$B=\frac{1}{2}(1-|\Phi|^2)$
is satisfied as a 
constraint, we deduce by theorem \ref{vortex},
that $\Psi_S(\tau)\in{\cal S}_N$, i.e.
the limit $\Psi_S(\tau)$ is a self-dual vortex for each 
$\tau\in[-\tau_*,\tau_*]$.
In addition,
by \eqref{h3} we have
$$
\|\Psi_\mu(\tau_1)-\Psi_\mu(\tau_2)\|_{H^1}\leq M_*|\tau_1-\tau_2|
$$
so that, by \eqref{convs}, the limit $\Psi_S$ will also satisfy
$$
\|\Psi_S(\tau_1)-\Psi_S(\tau_2)\|_{H^{r'}}\leq c|\tau_1-\tau_2|
$$
for $r'<1$, i.e. the limit is Lipschitz, and in particular lies in 
$W^{1,\infty}([-\tau_*,\tau_*];L^2)$.

For the second stage, we need to identify the limiting curve
$\tau\mapsto\Psi_S(\tau)\in{\cal S}_N$ as that described in
remark \ref{explicit}. It is clear, from the conditions on the initial
data in the statement of theorem \ref{main}, that 
$\Psi_S(0)=\psi_0\in{\cal S}_N$, and so it remains to deduce the 
ordinary differential equation \eqref{exh} which then determines
the curve completely.
To do this it is necessary to take the limit of \eqref{resceq}:
\beq\la{resceqmu}
\J\frac{\partial \Psi_\mu}{\partial \tau} = \mu \mcn ' + U ' + \J(dA_0^\mu ,
iA_0^\mu \Phi^\mu)
\eeq
 as 
$\mu\to\infty$. 
The first term on the right hand side is the most evidently
problematic. However, since the limiting motion is constrained to
the vortex space ${\cal S}_N$, it is only necessary to take
a limit projected onto the tangent space $T_{\Psi_S}{\cal S}_N$.
To this end, it is actually most convenient to introduce
$\proj_\mu(\tau)=\proj_{\Psi_{\mu}(\tau)}$ 
the spectral projection operator onto
$
\Ker\mcd_{\Psi_\mu(\tau)}=
\Ker\mcdadj_{\Psi_\mu(\tau)}\mcd_{\Psi_\mu(\tau)},
$ 
discussed
in lemma \ref{specproj}. By the final statement 
of lemma \ref{specproj},
and the convergence of $\Psi_{\mu_j}$ in \eqref{convs},
we know that $\proj_\mu(\tau)$ converge,
in the $L^2\to L^2$ operator norm, to the operator
$\proj_{\Psi_S(\tau)}$,
which is the spectral projection operator onto
$
\Ker\mcd_{\Psi_S(\tau)}=
\Ker\mcdadj_{\Psi_S(\tau)}\mcd_{\Psi_S(\tau)}.
$ 
(This latter operator is also the orthogonal $L^2$ projector
onto the tangent space $T_{\Psi_S}{\cal S}_N$ (subject to the
gauge condition \eqref{gauge1}).
Apply the operator $\proj_\mu(\tau)$ to the equation
\eqref{resceq}, to obtain:
\beq\la{projeq}
\proj_\mu(\tau)\J\frac{\partial \Psi_\mu}{\partial \tau} 
=\proj_\mu(\tau) U '(\Psi_\mu(\tau)),
\eeq
since $\J(dA_0 ,iA_0 \Phi_\mu)$ and $\mcn'(\Psi_\mu)$ 
are both in the kernel of
$\proj_\mu$ by lemma \ref{specproj}. We can now identify the
limit of the right hand side as $\proj_{\Psi_S(\tau)} U '(\Psi_S(\tau))$
at each $\tau$, and the convergence is strong in $L^2(\Sigma)$, by
\eqref{convs} and the above mentioned convergence of $\proj_\mu(\tau)$.
For the left hand side it is necessary to consider
the limit of the derivatives $\frac{\partial\Psi_\mu}{\partial\tau}$. 
Noting that these are bounded in e.g. $L^2([-\tau_*,\tau_*];L^2(\Sigma))$,
we may assume (by restricting to a further subsequence if necessary), 
the weak in $L^2$ subsequential convergence to a limit which is the weak time
derivative of $\Psi_S$:
$$
\langle \tilde f,\frac{\partial\Psi_{\mu_j}}{\partial\tau}
\rangle_{L^2([-\tau_*,\tau_*];L^2(\Sigma))}\to
\langle \tilde f,\frac{\partial\Psi_{S}}{\partial\tau}
\rangle_{L^2([-\tau_*,\tau_*];L^2(\Sigma))},
$$
for every $\tilde f\in L^2([-\tau_*,\tau_*];L^2(\Sigma))$. 
Now to identify the limit along
a convergent subsequence $\mu_j\to+\infty$,
consider the
projection operator $\proj_{\Psi_S(\tau)}$.
Choosing $\tilde f(\tau,\cdot)=\proj_{\Psi_S(\tau)}(f(\tau,\cdot))$, 
and using the symmetry
of  $\proj_{\Psi_S(\tau)}$ this implies that
\begin{align}
\int_{-\tau_*}^{+\tau_*}
\langle f,\,&\proj_{\Psi_S(\tau)}\J\frac{\partial\Psi_{\mu_j}}{\partial\tau}
\rangle_{L^2(\Sigma)}d\tau
=\int_{-\tau_*}^{+\tau_*}
\langle \proj_{\Psi_S(\tau)}f,\,\J\frac{\partial\Psi_{\mu_j}}{\partial\tau}
\rangle_{L^2(\Sigma)}d\tau\notag\\
&\to
\int_{-\tau_*}^{+\tau_*}
\langle \proj_{\Psi_S(\tau)}f,\,\J\frac{\partial\Psi_{S}}{\partial\tau}
\rangle_{L^2(\Sigma)}d\tau
=
\int_{-\tau_*}^{+\tau_*}
\langle f,\,\proj_{\Psi_S(\tau)}\J\frac{\partial\Psi_{S}}{\partial\tau}
\rangle_{L^2(\Sigma)}d\tau,\notag
\end{align}
for any $f\in L^2([-\tau_*,\tau_*];L^2(\Sigma))$. 
On the other hand, by the above mentioned convergence of
$\proj_\mu(\tau)$ 
to $\proj_{\Psi_S(\tau)}$ and the bounded convergence theorem
we have
$$
\int_{-\tau_*}^{+\tau_*}\Biggl[
\langle \proj_{\mu_j}(\tau)f,\J\frac{\partial\Psi_{\mu_j}}{\partial\tau}
\rangle_{L^2(\Sigma)}d\tau
-
\langle \proj_{\Psi_S(\tau)}f,\J\frac{\partial\Psi_{\mu_j}}{\partial\tau}
\rangle_{L^2(\Sigma)}\Biggr]d\tau\to 0,
$$
on account of the bound \eqref{h3}.
Therefore, we have in the limit:
\beq\la{projmod}
\int_{-\tau_*}^{+\tau_*}
\langle f,\proj_{\Psi_S(\tau)}\J\frac{\partial \Psi_S}{\partial \tau}
\rangle_{L^2(\Sigma)} d\tau
=\int_{-\tau_*}^{+\tau_*}
\langle f, \proj_{\Psi_S(\tau)} U '(\Psi_S(\tau))\rangle_{L^2(\Sigma)}
d\tau,
\eeq
for any $f\in L^2([-\tau_*,\tau_*];L^2(\Sigma))$. But since the limit is known
by the above to be in $W^{1,\infty}([-\tau_*,\tau_*];L^2)$, it is
differentiable (with respect to $\tau$, as a map into $L^2$) almost everywhere  (the standard result extends to Hilbert space-valued functions, see, e.g., \cite[prop. 6.41]{bl});  the derivative lies in the tangent space $T_{\Psi_S}{\cal S}_N$, which is
the range of the projector $\proj_{\Psi_S(\tau)}$. Consequently
\eqref{projmod} implies that $\tau\mapsto\Psi_S(\tau)$ is a solution
of \eqref{exh}, with equality holding in $L^2$ for almost every $\tau$.
But this in turn implies that $\tau\mapsto\Psi_S(\tau)$ is actually
continuously differentiable  into $L^2$, and we have a classical
solution of \eqref{exh}.

Finally for the third stage: we have now 
identified the limit as a solution of the limiting Hamiltonian
system specified using remark \ref{explicit}. Choosing smooth co-ordinates
on ${\cal M}_N$ as in \cite{s99} we see that this is a smooth finite dimensional Hamiltonian
system, and as such its solutions (for given initial data) are unique. 
Therefore
all subsequences have the same limit, and so we can assert full convergence
without resort to subsequences.

\section{Equations and identities related to the self-dual structure}
\la{idsd}
\begin{itemize}
\item[]
{\bf Notation change:}  {\em In this section time  does
not appear at all, and so the boldface
$\mbA$ for the spatial component is not used: i.e. in this section only,
$A$ refers to the spatial part of the connection, 
$A = A_1 dx^1+ A_2 dx^2$.}
\end{itemize}

Ginzburg-Landau vortices are critical points of the static Ginzburg
Landau energy functional $\mcn_{\lambda}=\int_\Sigma v_{\lambda}(A,\Phi)d\mu_g$
introduced following \eqref{glen}.
The {\em coupling}
constant $\lambda >0 $ is central to  the theory of critical points of the
Ginzburg-Landau 
functional   and the value $\lambda =1  $ is special as in this case 
the functional  admits the {\em Bogomolny decomposition}
introduced in remark \ref{bogom}. This allows for a detailed understanding
of the critical points not available for general values of $\lambda$,
and the theory of critical points for such general values is incomplete. (There is, however,
a substantial literature on the asymptotic behaviour of critical points
in the $\lambda\to+\infty$ limit, starting with \cite{br95}; 
see \cite{ss07} and references therein.)
This decomposition of  $\mcn\equiv\mcn_{1}$ has 
proved to be very useful not only for the analysis of critical points, 
but also for the 
associated time-dependent equations of vortex motion. For our purposes we need
in particular to derive a special form for the operator $L_\psi$ associated to the Hessian of
$\mcn$, see \eqref{hessdc}.
\subsection{Complex structure}
To discuss the Bogomolny structure in detail
it is useful to use a   complex formulation so we introduce the
complex co-ordinate
$z=x^1+ix^2$ for the complex structure $J$ on $\Sigma$.
Using this, there is a decomposition
of the complex 1-forms $\Omega^1_\C=\Omega^{{1,0}}\oplus\Omega^{{0,1}}$
into the $\pm i$ eigenspaces of $J$, see notation \ref{note}.  Let $\Omega^p(L)$ be the 
space of $p$-forms taking values in the bundle $L$:
then for $p=1$ there is a similar decomposition,
$$\Omega^1(L)=
\Omega^{{1,0}}(L)\oplus\Omega^{{0,1}}(L).
$$
Applying this decomposition to 
$D\Phi\in\Omega^1(L)$ we are led to introduce
the operator 
$D^{{0,1}}$ given by
$$D^{{0,1}}\Phi=\frac{1}{2}\Bigl((\nabla_1-iA_1)+i(\nabla_2-iA_2)\Bigr)\Phi
d{\overline z}\ = \ \barpa \Phi d\bar{z}.$$
For {\em real} 1-forms $A_1dx^1+A_2dx^2\in\Omega^1_\R$ 
this decomposition reads
$$
A_1dx^1+A_2dx^2=\alpha dz+\bar\alpha d\bar z,
$$
where $\alpha = \frac{A_1 -  iA_2}{2}$, and the map
$A\mapsto\alpha$ (resp. $A\mapsto\bar\alpha$) is an 
$\R$-linear isomorphism from $\Omega^1_\R$ to $\Omega^{{1,0}}$
(resp.  $\Omega^{{0,1}}$), and 
$\|A\|_{L^2}^2=4\int\bar\alpha\alpha e^{-2\rho}d\mu_g$.
With this $\alpha$ notation we can write
$$
\barpa\Phi = \frac{\partial\Phi}{\partial\bar{z}} - i\bar{\alpha}\Phi.
$$

\subsection{The Hessian}

The Bogomolny decomposition amounts to the
observation that, with $\lambda=1$,
$$
\mcn(A,\Phi)\equiv\mcn_{1} (A,\Phi) = \frac{1}{2}\int_\Sigma \Bigl(
4|\barpa \Phi|^2e^{-2\rho} \;+\;(B- \frac{1}{2}\left(1
  - |\Phi|^2)\right)^2\Bigr) \ \dmug  \ + \ \pi N
$$
where $N= deg L$.
If the following  first order equations, called the Bogomolny equations,
\begin{align}
\begin{split}
&\barpa\Phi=0,\\
B-&\frac{1}{2}(1-|\Phi|^2)=0\\
\end{split}
\la{bogo}
\end{align}
have solutions in a given class, they will automatically minimize
$\mcn$ within that class.

We introduce the nonlinear 
 Bogomolny operator associated to this decomposition,
\begin{align*}
\mcb  \;:\; \Omega^1_\R \oplus \Omega^0(L) 
  & \longrightarrow \Omega^0_\R \oplus \Omega^{0,1} (L)\\
(A,\Phi)
  & \mapsto \left(B- \frac{1}{2} (1 -|\Phi|^2) \ , \ \barpa \phi\right).
\end{align*}
Using the norm 
$\|(\beta,\eta)\|_{L^2}^2=\int (|\beta|^2+4e^{-2\rho}|\eta|^2)d\mu_g$ 
induced from the metric on the target
space, 
we see that 
$\mcn(A,\Phi)=\frac{1}{2}\|{\cal B}(A,\Phi)\|_{L^2}^2+\pi N$
as in remark \ref{bogom}; see \cite{brad88}.
The derivative of ${\cal B}$ at $\psi = (A, \Phi)$ is the map
$D\mcb_{\psi}  :  \Omega^{1}_{\Real} \oplus \Omega^0(L) 
\lra \Omega^0_\R \oplus \Omega^{0,1}(L)$ given by
\beq\la{map}
(\dot A, \dot \Phi) \ \mapsto \ (\ast d\dot A 
+\langle\Phi,\dot\Phi\rangle \,
, \, \barpa\dot\Phi - i\dot{\bar\alpha}\Phi)
\eeq
where $\alpha = \frac{A_1 -  iA_2}{2}$ and $\dot\alpha = \frac{\dot
  A_1 -i\dot A_2}{2}$.
Using this complex notation
allows a simple unified
formulation, which  takes account of the 
{\it gauge condition} \eqref{gauge1}:
this condition is the real part of
\begin{equation}
\label{gauge2}
4e^{-2\rho}\bar\partial\dot{\alpha} - i \Phi\bar{\dot\Phi} \ =\ 0,
\end{equation}
while the imaginary part of this expression 
is just the condition 
$\ast d\dot A +(\Phi,\dot\Phi)=0$, appearing in the 
linearized Bogomolny equations.
This suggests the introduction of the operators
\begin{align}
\begin{split}
\mcdpsi &: \left(\Omega^{1,0} \oplus \Omega^{0}(L)\right) \lra
\left(\Omega^{0}_\C\oplus\Omega^{0,1}(L)\right) \\
\mcdpsi^{\ast}
  &:\left(\Omega^{0}_\C \oplus \Omega^{0,1}(L)\right) \lra
\left(\Omega^{1,0}\oplus\Omega^{0}(L)\right)
\label{delta}
\end{split}
\end{align}
given by 
\begin{align}
\begin{split}
\mcdpsi (\dot{\alpha}, \dot{\Phi}) & = (4e^{-2\rho}\bar{\partial}\dot{\alpha} - i\Phi\bar{\dot{\Phi}}, \ \barpa\dot{\Phi} - i\bar{\dot{\alpha}}\Phi)\\
\mcdpsi^{\ast} (\beta, \eta) & = (-\partial\beta - i\Phi\bar\eta, \
-4 e^{-2\rho} \pa\eta - i\Phi\bar{\beta}).
\end{split}
\end{align}
We use the real inner product associated to
the $L^2$ norms induced from the metric as above, i.e.:
\begin{align}
\Bigl\langle(\dot\alpha,\dot\Phi),(\alpha',\Phi')\Bigr\rangle_{L^2}
&=\int \bigl(4\er\Re{\bar{\dot\alpha}}\alpha'+\Re\bar{\dot\Phi}\Phi'\bigr)\, \dmug
\qquad\hbox{on}\;\Omega^{1,0} \oplus \Omega^{0}(L)\notag\\
\Bigl\langle(\beta,\eta),(\beta',\eta')\Bigr\rangle_{L^2}
&=\int \bigl(\Re\bar\beta\beta'+4\er\Re\bar\eta\eta'\bigr)\, \dmug\qquad\;\hbox{on}\;
\Omega^{0}_\C\oplus\Omega^{0,1}(L)
\;
.\notag
\end{align}
Integrating by parts we deduce that
$$
\Bigl\langle\mcdpsi(\dot\alpha,\dot\Phi),(\beta,\eta)\Bigr\rangle_{L^2}=
\Bigl\langle(\dot\alpha,\dot\Phi),\mcdpsi^{\ast}(\beta,\eta)\Bigr\rangle_{L^2}
$$
so that $\mcdpsi^{\ast}$ is the $L^2$ adjoint of $\mcdpsi$ and 
\begin{align}
 \la{dstard}
\mcdpsi^{\ast}\mcdpsi (\dot\alpha, \dot\Phi)& = \Big(-\partial ( 4\er\bar{\partial}\dot{\alpha}- i
  \Phi\bar{\dot{\Phi}} )  - i \Phi
  (\overline{\barpa\dot{\Phi}}+ i\dot{\alpha}\bar\Phi)\ ,\\
 &\qquad\qquad - 4\er\pa(\barpa\dot\Phi - i\bar{\dot\alpha}\Phi) -
  i\Phi( 4\er\partial\bar{\dot\alpha} + i\bar\Phi\dot{\Phi})\Big)\notag\\
& =\ \Big(-\partial( 4\er\bar{\partial}\dot{\alpha}) +i
(\pa\Phi)\bar{\dot\Phi}+|\Phi|^2\dot{\alpha}\ , \  -4\er\pa\barpa\dot{\Phi} +
|\Phi|^2\dot\Phi + i 4\er\bar{\dot\alpha}\pa\Phi\Big).\notag
\end{align}
We compare this expression with the operator defined in 
\eqref{modhess}:
\begin{equation}
{\overline L}_\psi \ :\  \left(\Omega^{1,0} \oplus \Omega^{0}(L)\right) 
\lra \left(\Omega^{1,0} \oplus \Omega^{0}(L)\right) 
\end{equation}
which defines the Hessian of $\mcn$ on the subspace on which the gauge
condition \eqref{gauge1} is satisfied,
i.e.,
\begin{equation}
\langle \dot\psi ,{\overline L}_\psi \dot \psi \rangle_{L^2}  = D^2\mcn_\psi (\dot\psi,\dot\psi) = \frac{d^2}{d\epsilon^2}\vert_{\epsilon =0} \mcn (\psi +\epsilon \dot \psi),
\end{equation}
for $\dot\psi = (\dot A, \dot \Phi)$ satisfying 
\eqref{gauge1}.
Using mixed real/complex notation for $A/\alpha$, \eqref{modhess}
implies the following formula:
\begin{align}\la{hesssd}
{\overline L}_\psi = \Big( -4\partial(\er\bar\partial \dot\alpha) 
&+|\Phi|^2\dot\alpha  -
(i\dot\Phi , D_1 \Phi) + i (i\dot\Phi, D_2\Phi)
\ \ , \notag\\ 
&-\Delta_A\dot\Phi -\frac{1}{2} (1 - 3 |\Phi|^2)\dot
\Phi +2i\er\dot{A} \cdot D\Phi\Big).
\end{align}
Calculate $\dot A\cdot D\Phi=2\dot\alpha\barpa\Phi
+2\bar{\dot\alpha}\pa\Phi$ and
$ -(i\dot\Phi , D_1 \Phi) + i (i\dot\Phi, D_2\Phi)
=i\bar{\dot\Phi}\pa\Phi-i\dot\Phi\overline{\barpa\Phi}$,
from which it follows that 
\beq
\la{hessdc}
({\overline L}_\psi  - \mcdpsi^{\ast}\mcdpsi ) \dot \psi = 
               \left(
               \begin{array}{l}
               -i\dot \Phi \overline{\barpa{\Phi}}\\
               \left(B- \frac{1}{2}(1 - |\Phi|^2)\right)\dot\Phi + 4i\er
               \dot\alpha\barpa\Phi 
               \end{array}
               \right).
\eeq
(Incidentally, observing that
$$
{\cal B}(A+\dot A,\Phi+\dot\Phi)=
{\cal B}(A,\Phi)+\mcdpsi\dot\psi+
\left(\frac{1}{2}|\dot\Phi|^2,-i\dot\bar\alpha\dot\Phi\right),
$$
with $\dot\psi = (\dot A, \dot \Phi)$ satisfying 
\eqref{gauge1}, the identity \eqref{hessdc}
can also  be read off from the quadratic part of the Taylor expansion
for $\mcn(A+\dot A,\Phi+\dot\Phi)$:
\begin{align*}
\frac{1}{2}\langle \dot\psi ,{\overline L}_\psi \dot \psi \rangle_{L^2}
  & = \onetwo|{\cal D}_\psi \dot
     \psi|^{2}_{L^2}  + \left\langle\mcb(\psi) \ ,  \left(\onetwo
         |\dot\Phi|^2, -
         i\dot{\bar\alpha}\dot\Phi\right)\right\rangle\\
  & = \onetwo |{\cal D}_\psi \dot
     \psi|^{2}_{L^2} + \int_\Sigma  \left( \onetwo (B-\onetwo(1
       -|\Phi|^2))|\dot\Phi|^2 + 4\er\langle\barpaphi ,
       -i\dot{\bar\alpha}\dot\Phi\rangle\right) \ \dmug  ,
\end{align*}
using the inner product on $\Omega^{1,0} \oplus \Omega^{0}(L)$
defined above.)
\begin{corollary}
\la{c31}
Let $\J$ denote the complex structure defined in \eqref{defcom}.
There exists a number $c>0$, independent of  
$\psi=(\alpha,\Phi)$ and $\zeta=\dot\psi=(\dot\alpha,\dot\Phi)
\in\Omega^{1,0} \oplus \Omega^{0}(L)$, such that  
$$
|\langle  \J\zeta, L_\psi
\zeta\rangle_{L^2} | \ \leq c\  |\mcb(\psi)|_{L^\infty}|\zeta|^{2}_{L^2}
$$
\end{corollary}
\proof  
By \eqref{hessdc}
$ |\langle  \J\zeta, L_\psi
\zeta\rangle_{L^2}  - \langle \mcdpsi \J\zeta ,\mcdpsi\zeta\rangle_{L^2}|
\leq |\mcb(\psi)|_{L^\infty}|\zeta|^{2}_{L^2}.   
$
Now the complex structure $\J$ written in complex notation,
i.e. acting on $\Omega^{1,0} \oplus \Omega^{0}_(L)$, is given by
$\J(\dot\alpha,\dot\Phi)=(-i\dot\alpha,i\dot\Phi)$. Correspondingly,
on $ \Omega^{0}_\C\oplus\Omega^{0,1}(L)$ we introduce the complex structure
$\J'(\beta,\eta)=(i\beta,-i\eta)$. Then,
by observation 
\beq\la{jlin}
\mcdpsi \J\zeta = -\J'\mcdpsi \zeta.
\eeq
Therefore,
writing $w= \mcdpsi\zeta$, we have 
$\langle \mcdpsi \J\zeta , \mcd_\psi\zeta\rangle_{L^2} = \langle
-\J' w, w\rangle_{L^2} = 0$ by skew-symmetry,
and the result follows.\qed

\begin{lemma}
\la{bdq}
Assume there are positive numbers $L,\mce_0$ such that 
$|\Phi|_{L^2} = L$, and $\mcn_\lambda(A,\Phi)=\mce_0$
and $\lambda>0$. Then
the quadratic forms
\begin{align*}
  &\tilde{Q}_{\Phi} (\beta) = \int_\Sigma 4|\partial \beta|^2 \er + |\Phi|^2|\beta|^2 \dmug\;\hbox{\ \ on }\;\oplus \Omega^{0}_\C
\hbox{and} \\
  &Q_{(A,\Phi) } (\eta) = \int_\Sigma  4 e^{-4\rho} |\pa \eta|^2 +\er |\Phi|^2 |\eta|^2 \dmug\;\hbox{\ \ on }\;\Omega^{0,1}(L)  
\end{align*}
are strictly positive, and in fact bounded below by (respectively) 
$C\|\beta\|_{H^1}^2$ and $C\|\eta\|_{\honea}^2$
where $C$ is a positive number depending only upon the numbers
$L,\mce_0$.
\end{lemma}
\proof
We will present the proof for the quadratic form 
$
Q_{(A,\Phi)}(\eta)
$
as the other is similar but easier.
Clearly $Q_{(A,\Phi)}(\eta) \geq 0$ and in fact $Q_{(A,\Phi)}(\eta) =0$ if and only if $\eta\equiv 0$ on $\Sigma$ (because if $\pa\eta \equiv 0 $ then $\eta$ has isolated zeros (as in \cite{jt82}, sec. 3.5);  if $\Phi\eta\equiv 0$ then $\eta \equiv 0$ since $\Phi = 0 \ a.e.$  contradicts $\int_\Sigma |\Phi|^2 = L>0$. 
Furthermore, we show that 
$
Q_{(A,\Phi)}(\eta) \geq c |\eta|_{L^2}^{2}
$
for a constant $c$; to be precise
there exists $c=c(L,\mce_0)$ such that 
\beq\la{pbc}
Q_{(A,\Phi)}(\eta) \geq c,\;\hbox{for all $\eta$ such that}\,\|\eta\|_{L^2}=1.
\eeq
We will prove this by contradiction. First we obtain some bounds.
By gauge invariance we are free to assume
that the Coulomb gauge condition $\dv A=0$ holds.
With this gauge condition, we have the bound
$\|A\|_{H^1}\le c(\mce_0)$ and so $A$ is bounded in every $L^p$ space.
Now use $\|\partial\eta\|_{L^p}\leq\|\pa\eta\|_{L^p}+\|A\eta\|_{L^p}$
to deduce that
$$\|\partial\eta\|_{L^p}^2\leq C(1+Q_{(A,\Phi)}(\eta))$$ 
for every $p<2$,
by Holder's inequality. This in turn implies,
by the $L^p$ estimate for the inhomogeneous
Cauchy-Riemann system,
that $\eta$ is bounded similarly
in $L^4$, and so since $A$ is also we can bound $\partial\eta$ in $L^2$ 
and hence $\eta$ in $H^1$.
Finally, since $A$ and $\eta$ are bounded similarly
in $L^4$, this imples that
$\|\eta\|^2_{\honea}\leq C(1+Q_{(A,\Phi)}(\eta))$, with
$C$ depending only upon $\mce_0,L$.
To conclude, in Coulomb gauge the $A,\Phi,\eta$ are all bounded
in $H^1$ in terms of $L,\mce_0,Q_{(A,\Phi)}(\eta)$.

The contradiction argument now starts: assume \eqref{pbc} fails.
Then, by the bounds just obtained and the Banach-Alaoglu and Rellich theorems, 
there is a sequence $(A_\nu, \Phi_\nu, \eta_\nu)$ with 
$$\|A_\nu\|_{H^1}+\|\nabla\Phi_\nu\|_{L^2} \leq K(\mce_0,L),$$  
$\|\Phi_\nu\|_{L^2}= L$ and $\|\eta_\nu\|_{L^2} =1$, such that 
\begin{align*}
& Q_{(A_\nu,\Phi_\nu)}(\eta_\nu)\lra 0\\
& A_\nu  \lra A \ \ \mbox{weakly in $H^1$}\\
& \Phi_\nu \lra \Phi \ \ \mbox{weakly in $H^1$ and strongly in $L^p$ for any $p<\infty$}\\
& \eta_\nu \lra \eta \ \  \mbox{weakly in $H^1$ and strongly in $L^p$.}
\end{align*}
This implies that 
$|\Phi|_{L^2} = L>0,\, Q_{(A,\Phi)}(\eta) = 0 $
which implies as above that $\Phi = 0  \ a.e. $ and contradicts as above that
$|\Phi|_{L^2} $ is constant. 
This leads to
$$
Q_{(A,\Phi)}(\eta)\geq c_1 |\eta|_{L^2}^{2}\ \mbox{where}\ c_1= c_1(L,\mce_0).
$$
Finally just apply the bound above for $\|D\eta\|_{L^2}$
to improve this up to the $H^1_A$ lower bound claimed.\qed

\subsection{The Bogomolny foliation}

We introduce a foliation associated to the Bogomolny operator, which
we regard as a map between the following Hilbert spaces:
\begin{align*}
\mcb  \;:\; H^1\bigl(\Omega^1_\R \oplus \Omega^0(L) \bigr)
& \longrightarrow L^2\bigl(\Omega^0_\R \oplus \Omega^{0,1} (L)\bigr),\\
(A,\Phi)
  & \mapsto \left(B- \frac{1}{2} (1 -|\Phi|^2) \ , \ \barpa \phi\right).
\end{align*}
With this choice of norms $\mcb$ is a smooth function. The next result
shows that it is a submersion if the energy is close to the minimum value:
\begin{lemma}
\la{w}
There exists $\theta_*>0$ such that 
$\|{\barpa{\Phi}}\|_{L^2}<\theta_*$ implies that 
$\Ker\mcdadj_{\Psi}=\{0\}$, and $\Ker\mcd_{\Psi}$ is $2N$
dimensional (where $N= deg L$).

\end{lemma}
\proof
$\mcdadj (\beta, \eta) = 0 $ is equivalent to 
\begin{align*}
  & -\partial \beta - i \Phi\bar\eta = 0 \\
  & - 4 \er \pa \eta - i \Phi \bar\beta = 0.
\end{align*}
Apply the operations $4\bar\partial$ to the first and $4 \barpa$ to
the second of these equations to deduce that 
\begin{align*}
  & -4\er \bar\partial\partial\beta +|\Phi|^2 \beta - 4 i\er \barpa \Phi \bar\eta = 0 \\
  & -4 \barpa (\er \barpa \eta) + |\Phi|^2\eta - i (\barpa \Phi)\beta = 0
\end{align*}
The first two terms of these two equations are respectively the Euler-
Lagrange operators associated to the quadratic forms
$\tilde{Q}_{\Phi}(\beta)$ and
$Q_{A,\Phi } (\eta)$ studied in the previous lemma.
Then we get the estimates 
\begin{align*}
\tilde Q_\Phi (\beta) &\leq c |\barpa \Phi|_\ltwo |\beta|_\lfour |\eta|_\lfour \\\
Q_{A,\Phi}  (\eta) & \leq c |\barpa \Phi|_\ltwo |\beta|_\lfour |\eta|_\lfour 
\end{align*}
which implies the result, since 
$
\tilde{Q}_\Phi (\beta)  \ \geq \ c|\beta|^{2}_{\hone} 
$
and 
$
Q_{A,\Phi} (\eta) \ \geq \ c|\eta|^{2}_{\honea}  .
$\qed

The natural
geometrical context
for the results of this section will now be explained.
Define
$
{\cal O}_*\equiv
\{
(A,\Phi)\in H^1(\Omega^1_\R \oplus \Omega^0(L)):
\|\barpa{\Phi}\|_{L^2}<\theta_*\}
$
which is an open set containing 
$\{\psi = (A, \Phi):\mcb(\psi)=0\}\subset
H^1(\Omega^1_\R \oplus \Omega^0(L))$.
Furthermore, the previous lemma implies that
$\mcdpsi:(\Omega^{1,0} \oplus \Omega^{0}(L)) \lra
(\Omega^{0}_\C\oplus\Omega^{0,1}(L))$ 
is surjective for $\psi\in{\cal O}_*$. By the discussion
in the paragraph preceding \eqref{delta}, this implies
that $D\mcb_{\psi}  :  \Omega^{1}_{\Real} 
\oplus \Omega^0(L) 
\lra \Omega^0_\R \oplus \Omega^{0,1}(L)$
is also surjective for $\psi\in{\cal O}_*$, 
and hence the level sets of
$\mcb$ form a foliation of ${\cal O}_*$ whose leaves 
have tangent space equal to 
$\Ker D\mcb_{\psi}$ by
\cite[\S 3.5 and \S 4.4]{amt}. The intersection of this tangent 
space with ${\cal SL}_\psi=\{(\dot A,\dot\Phi):
(\dot A,\dot\Phi)$ satisfies
\eqref{gauge1}$\}$ is  $\Ker\mcd_{\Psi}$.

%
\begin{lemma}
\label{specproj}
Assume $\psi\in (\Omega^{1,0} \oplus \Omega^{0}_\C(L))\cap{\cal O}_*$.
The operators $\mcdadjp\mcdpsi$ defined in \eqref{dstard} are 
self-adjoint operators on $L^2$, with domain $H^2$, 
with $2N$-dimensional kernel equal to 
$\Ker\mcdpsi$, and:
\beq
\la{coe}
\|\mcdadjp\mcdpsi\zeta\|_{L^2}+\|\zeta\|_{L^2}\geq c\|\zeta\|_{H^2}.
\eeq
Let $\proj_{\psi}$ be the orthogonal spectral
projector onto $\Ker\mcdadjp\mcdpsi=\Ker\mcdpsi$. 
Then $\proj_{\psi}(\mcn'(\psi))=0$ and 
$\proj_{\psi}(\J(d\chi ,i\chi \Phi_\mu))=0$ for any smooth real
valued function $\chi$.
Finally, if also 
$\psi^{(j)}\in (\Omega^{1,0} \oplus \Omega^{0}_\C(L))\cap{\cal O}_*$, 
and
$\sup_j\|\psi^{(j)}\|_{\ch_2}<\infty$ and 
$\lim_{j\to +\infty}\|\psi^{(j)}-\psi\|_{\ch_r}=0$, for 
all $r<2$, the corresponding
projectors $\proj_{\psi^{(j)}}$ converge to $\proj_{\psi}$ in 
$L^2\to L^2$ operator norm.
\end{lemma}
\proof
The first assertion and the bound \eqref{coe} follow from
lemma \ref{w} and standard elliptic theory. The next statement
follows by noting that if $n\in\Ker\mcdpsi$, then 
differentiation of 
$\mcn(\psi)=\frac{1}{2}\int|{\cal B}(\psi)|^2d\mu_g+\pi N$ yields
$$
\langle n,\mcn'(\psi)\rangle_{L^2}=\frac{d}{ds}\bigg|_{s=0}\mcn(\psi+s n)
=\langle \mcb(\psi),D\mcb_\psi(n)\rangle_{L^2}=0
$$
since $\Ker\mcdpsi\subset \Ker D\mcb_\psi$ by the discussion
preceding \eqref{delta}. Next,
$n\in\Ker\mcdpsi$ implies that $\proj_{\psi}
(\J(d\chi ,i\chi \Phi_\mu))=0$ since  integration by parts 
reduces this to the fact that $n$ solves the first component
of $D\mcb_\psi n=0$ in \eqref{map}.

The final statement follows by \cite[\S\,IV.3]{kato80}, if it can
be established that
$T_j\equiv\mcdadj_{\psi^{(j)}}\mcd_{\psi^{(j)}}$ converges to 
$T\equiv\mcdadjp\mcdpsi$ in the generalized sense of Kato
(see \cite[\S IV.2.6]{kato80}), or equivalently in the norm
resolvent sense:
\beq\la{rs}
\lim_{j\to\infty}\|(i+T)^{-1}-(i+T_j)^{-1}\|_{L^2\to L^2}=0.
\eeq
To verify this convergence, it is convenient first of all to verify
it in Coulomb gauge. So let 
$
\tilde\psi^{(j)}=(\tilde A^{(j)},\tilde\Phi^{(j)})=e^{i\chi_j}\cdot\psi^{(j)}
$
and  
$\tilde\psi=(\tilde A,\tilde\Phi)=e^{i\chi}\cdot\psi$ 
be gauge transforms
(as defined following \eqref{gg}), 
such that $\dv\tilde A^{(j)}=0=\dv\tilde A$. The assumed properties of
$\psi^{(j)}$ ensure that $\sup\|\chi_j\|_{H^2}<\infty$ and that
$\lim\|\chi_j-\chi\|_{H^r}=0,\,\forall\, r<2$ so that also
$\tilde\psi^{(j)}\to\tilde\psi$ in $\ch_r$ for $r<2$.
Now observe that {\em in Coulomb gauge} the formula \eqref{dstard} does not
involve any derivatives of the connection one-form $A$ at all. From
this it is then immediate by inspection that (writing
$\tilde T_j\equiv\mcdadj_{\tilde\psi^{(j)}}\mcd_{\tilde\psi^{(j)}},
\;\hbox{and}$
$\tilde T\equiv\mcdadj_{\tilde\psi}\mcd_{\tilde\psi},$)
\beq
\la{gk}
\|(\tilde T-\tilde T_j)\zeta\|_{L^2}\leq 
\delta_j\|\zeta\|_{H^2}\leq
c\delta_j(\|\zeta\|_{L^2}+
\|\tilde T\zeta\|_{L^2})
\eeq
where $\delta_j\to 0$ as $j\to+\infty$. But this last fact implies
(by \cite[Theorems IV.2.24-25]{kato80}) that $\tilde T_j$ converges
to $\tilde T$ in the generalized sense, and hence in the resolvent sense:
\beq\la{trs}
\lim_{j\to\infty}\|(i+\tilde T)^{-1}-(i+\tilde T_j)^{-1}\|_{L^2\to L^2}=0.
\eeq
This would establish the convergence of the corresponding spectral projectors
in Coulomb gauge. To go back to the original $\psi_j$ it is just necessary
to make use of the following gauge invariance property: on
$\zeta=(\dot\alpha,\dot\Phi)$ the induced action of the gauge group is
$g\bullet(\dot\alpha,\dot\Phi)=(\dot\alpha,g\dot\Phi)$ for any
$S^1$ valued function $g$, and
$$
\tilde T \left(e^{i\chi}\bullet\zeta\right)
=e^{i\chi} \bullet \left(T\zeta\right),
$$
and similarly with $T_j,\chi_j$ replaced by $T,\chi$.
This gauge invariance property implies that
$(i+T_j)^{-1}=e^{-i\chi_j}\circ(i+\tilde T_j)^{-1}\circ e^{i\chi_j}$
and $(i+T)^{-1}=e^{-i\chi}\circ(i+\tilde T)^{-1}\circ e^{i\chi}$, 
where by $\circ$ we mean operator composition, and $e^{i\chi}$ is
shorthand for the operator $e^{i\chi}\bullet$ etc. Finally,
using $\lim\|\chi_j-\chi\|_{H^r}=0,\,\forall\, r<2$ we see that
\eqref{gk} and \eqref{trs} imply \eqref{rs}, completing the proof.
\myqed
\section*{Appendix}
\setcounter{section}{1}
\setcounter{equation}{0}
\setcounter{subsection}{0}
\protect\renewcommand{\thesection}{\Alph{section}}
\label{app}
\subsection{Operators}

To describe in detail the Laplacian operators which appear in the
text, we assume $\Sigma$ to be covered by an atlas of charts
$U_\alpha$ on each of which is a local
trivialisation of $L$ determined by a choice 
of a local unitary frame.
(A smooth section $\Phi$ of $L$ then corresponds to a family of 
smooth functions $\Phi_\alpha:U_\alpha\to{\Complex}$
so that on
$U_\alpha \cap U_\beta$ we have  $\Phi_\alpha = e^{i\theta_{\alpha\beta}} 
\Phi_\beta$ with $e^{i\theta_{\alpha\beta}}:U_\alpha \cap
U_\beta \to S^1$ smooth.) 
We assume given a smooth connection ${\bf D}=\nabla-i\mbA$ on $L$ acting
as a covariant derivative operator on sections of $L$.
Working in such a chart, and suppressing the index $\alpha$,
the Laplacian on sections $\Phi$ of $L$ is given by:
\begin{equation}
-\Delta_A\Phi=-\frac{1}{\sqrt{g}}D_j\bigl(g^{ij}\sqrt{g}D_i\Phi\bigr)
=-e^{-2\rho}\bigl(D_i D_i\Phi\bigr).
\label{deflap}
\end{equation}
This satisfies $\langle-\Delta_A\Phi,\Phi'\rangle_{L^2}
=\frac{d}{d\epsilon} 
\frac{1}{2}|{\bf D}(\Phi+\epsilon\Phi')|_{L^2}^2|_{\epsilon=0}.
$

Next we need the Laplacian on one-forms. Starting with 
$\mbA=A_1dx^1+A_2dx^2\in\Omega^1_\R$, the negative Laplacian
is the Euler-Lagrange operator associated to the
Dirichlet form $\frac{1}{2}\int(|\dv \mbA|^2
+|\mathbf{d}\mbA|^2)d\mu_g$ (with the
norms inside the integral determined by $g$ in the standard way).
Transferring
to complex form $\alpha=\frac{1}{2}(A_1-iA_2)\in\Omega^{1,0}$, this
Dirichlet form is just 
$I(\alpha)=8\int e^{-4\rho}\overline{\bar\partial\alpha}
\bar\partial\alpha\, d\mu_g$.
The corresponding negative Laplacian $-\Delta^{1,0}$ is then
defined by $\langle -\Delta^{1,0}\alpha,\beta\rangle_{L^2}
=\frac{d}{d\epsilon} I(\alpha+\epsilon\beta)|_{\epsilon=0}$
where we use the induced inner product $\Omega^{1,0}$ as
in \S\ref{idsd}. This
leads to the following formula for the negative Laplacian 
$-\Delta^{1,0}$ on $\alpha\in\Omega^{1,0}$:
$$
-\Delta^{1,0}\alpha=-4\partial(e^{-2\rho}\bar\partial\alpha),
$$
which is precisely the operator appearing in \S\ref{idsd}.
Similarly, on $\Omega^{0,1}(L)$ the negative Laplacian is
$$
-\Delta_A^{0,1}\eta=-4\barpa(e^{-2\rho}\pa\eta),
$$
which is the operator in \eqref{eqeta}.

\subsection{ Norms and inequalities}

We define the Sobolev norms defined with the covariant
derivative ${\bf D}=\nabla_\mbA=\nabla-i\mbA$. (We write
$\nabla_\mbA$ in place of ${\bf D}$ for emphasis here.)
The first Sobolev norm is defined by
\beq
\la{dhonea}
|\Phi|^{2}_{\hoa} = 
\int_\Sigma\left( |\Phi|^2 + |\nabla_A \Phi|^2 \right)d\mu_g . 
\eeq
In the above integral  the inner products are the standard ones
induced from $h$ and $g$. The higher norms $\hta,\dots$ are 
defined similarly, as are the $W^{k,p}_\mbA$ norms for integral $k$
and any $p\in[1,\infty]$.
The $L^p$ norms of the higher covariant derivatives arising from
the connections $\nabla_\mbA$ and $\nabla$ are related as 
expressed schematically in the following:
\ba
\|\nabla\Phi\|_{L^p}&\le &  \|\nabla_\mbA\Phi\|_{L^p}
+c\|\mbA\|_{L^\infty}\|\Phi\|_{L^p},\la{end1}\\
\|\nabla\nabla\Phi\|_{L^p}&\le & \|\nabla_\mbA\nabla_\mbA\Phi\|_{L^p}
+c\|\mbA\|_{L^\infty}\|\nabla_\mbA\Phi\|_{L^p}\la{end2}\\
&&\qquad+
c(1+\|\nabla \mbA\ \Phi\|_{L^p}+\|\mbA\|^2_{L^{\infty}}
\|\Phi\|_{L^p}),
\notag\\
\|\nabla\nabla\nabla\Phi\|_{L^p}
&\le & \|\nabla_\mbA\nabla_\mbA\nabla_\mbA\Phi\|_{L^p} 
+c\|\mbA\|_{L^\infty}\|\nabla_\mbA\nabla_\mbA\Phi\|_{L^p}\la{end3}\\
&&\;+c(1+\|\nabla \mbA\|_{L^\infty}+\|\mbA\|^2_{L^{\infty}})
\|\nabla_\mbA\Phi\|_{L^p}\notag\\
&&\qquad +c\bigl(1+\|\nabla^2 \mbA\|_{L^q}\|\Phi\|_{L^r}+\|\mbA\|^3_{L^{\infty}}\|\Phi\|_{L^p}\bigr),
\notag
\ea
where $q^{-1}+r^{-1}=p^{-1}$.

We now collect together some inequalities from \cite{ds07}.

The system of equations
\begin{equation}
B=f\qquad \dv \mbA=g
\end{equation}
(where as above $\dv:\Omega^1\to\Omega^0$ is minus the adjoint of $d$)
is a first order elliptic system which can be solved for $\mbA$ subject 
to the condition on 
$\int f d\mu_g$ dictated by an integer $N$, the degree of $L$. 
It can be rewritten
\begin{equation}\label{hodgepodge}
\mathbf{d}\mbA=(f-b)d\mu_g\qquad \dv \mbA=g
\end{equation}
and solved via Hodge decomposition as long as the right hand sides have
zero integral. There is a solution unique up to addition of 
harmonic 1-forms which
satisfies $\|\mbA\|_{W^{1,p}}\le c_p(1+\|f\|_{L^p}+\|g\|_{L^p})$ 
for $p<\infty$.

\begin{lemma}[Covariant Sobolev and Gagliardo-Nirenberg inequalities]
\label{lemcovgn}
For \\\noindent
$(\Sigma, g)$  as  above and for  $(\mbA,\Phi)\in (H^1\times
H^{2}_{\mbA})(\Sigma)$ then $\nabla_\mbA\Phi\in L^4 (\Sigma)$ and 
\begin{align}
\|\naf\|_{L^4} 
  & \leq c \|\nabla_\mbA \Phi\|_{\hoa}\label{covsob}\\
\intertext{and also for all $1\le p<\infty$,
  $H^{2}_{\mbA}\hookrightarrow W^{1,p}_{\mbA} 
  \hookrightarrow \linf$ continuously on   $\Sigma$. Also}
\|\naf\|_{L^4}
 & \leq c \|\naf\|^{1/2}\sublt\Bigl(\|\naf\|^{1/2}\sublt
  +\|\nabla_\mbA \naf\|^{1/2}\sublt\Bigr) \label{covgn}
\end{align}
where $c$ depends only on $(\Sigma, g)$.
\end{lemma}
\begin{lemma}[Covariant version of the Garding inequality]
\label{lemgarding}
For  $\Psi = (\mbA,\Phi)$ such that the norms on
$\Sigma$ appearing  below are finite we have  
\begin{equation}
\label{gardeq}
\|\nabla_\mbA\nabla_\mbA\Phi\|_{L^2 } \leq 
\|\Delta_\mbA\Phi\|_{L^2}
+ c\|B\|_\linf^{1/2}\|\nabla_\mbA\Phi\|\sublt + 
  c \|\Phi \|^{1/2}_\linf\|\nabla_\mbA\Phi\|\sublt^{1/2}
\|\nabla B\|^{1/2}\sublt 
\end{equation}
where $c$ is a number depending only on $(\Sigma, g)$.
\end{lemma}
\begin{lemma}[Covariant version of the Brezis-Gallouet inequality]
\label{lemcovbg}
If $\mbA\in H^1(\Sigma)$  and $\Phi\in
\hta(\Sigma)$  then  
\begin{equation}
\label{covbg}
\|\Phi\|_{L^\infty (\Sigma)}\leq  c\Bigl( 1+ \|\Phi\|_{\hoa}\sqrt{\ln 
   ( 1+  \|\Phi\|_{H^{2}_{\mbA}}}) \Bigr)
 \end{equation}
where $c$ depends only on $(\Sigma , g)$.
\end{lemma}
\subsection{Global existence results and different choices of gauge}
\la{brez}
  
In this section we will summarize the existence 
theory for \eqref{scs} from \cite{bbs} and \cite{ds07},
and explain how theorem \ref{thglob2} can be deduced from it.
Existence theory can be worked out using various gauge conditions, and a choice of gauge is usually made to facilitate the calculations. 
The simplest condition for the statement of the theorem, 
which also is convenient if we wish to make the
Hamiltonian structure manifest - see \S\ref{varham},
is the temporal gauge condition $A_0=0$; however, the regularity is stronger in Coulomb gauge  $\dv \mbA =0$.  We have the following statements.  

\begin{theorem}[Global existence in temporal gauge]\label{thglob} 
Given data
$\Phi(0)\in H^2 (\Sigma)$ and  ${\bf A}(0) \in H^1(\Sigma)$,
there exists a global solution for  the Cauchy problem for \eqref{scs} 
satisfying $A_0=0$, with regularity
$\Phi\in  C\bigl([0,\infty); H^2 (\Sigma)\bigr)
\cap C^1\bigl([0,\infty); L^2 (\Sigma)\bigr)$
and ${\mathbf A}\in  C^1\bigl([0,\infty); H^1 (\Sigma)\bigr)$. 
Furthermore, it
is the unique such solution satisfying $A_0 =0$ and satisfies the 
estimate 
$$
\|\Phi(t)\|_{H^2(\Sigma)}\le ce^{\alpha e^{\beta t}},
$$
for some positive  constants $c, \alpha , \beta$ depending only on
$(\Sigma,g)$, the equations, and the initial data.  
\end{theorem}

This can be derived from theorem 1.1 in \cite{ds07}, by applying
a gauge transformation to put the solution obtained there into 
temporal gauge. To be precise the cited result gives a global solution
$(a_0,{\bf a},\phi)$ of the system \eqref{scs} satisfying the parabolic
gauge condition $a_0=\dv{\bf a}$, and the gauge invariant growth
estimate
\beq\la{gig}
\|\phi\|_{H^2_{a}(\Sigma)}(t)\leq ce^{\alpha e^{\beta t}}.
\eeq
The solution satisfies
$\phi\in  C\bigl([0,\infty); H^2 (\Sigma)\bigr)
\cap C^1\bigl([0,\infty); L^2 (\Sigma)\bigr)$,
${\bf a}\in  C\bigl([0,\infty); H^1(\Sigma)\bigr)$ and
${a_0}\in  C\bigl([0,\infty); L^2(\Sigma)\bigr)$. Now define
$\chi\in C^1\bigl([0,\infty); L^2(\Sigma)\bigr)$ by
$\partial_t  \chi +a_0=0$ and $\chi(0)=0$.
Define $({\Phi},{A})=(\phi e^{it\chi}, a+d\chi)$:
this gives a solution to \eqref{scs} satisfying the properties 
asserted in theorem \ref{thglob}. (Most of this can be read off immediately, except
perhaps to verify that ${\mathbf A}\in  C^1\bigl([0,\infty); H^1 (\Sigma)\bigr)$,  but this
follows from the first equation in \eqref{scs}, using the fact that $A_0=0$ and
the right hand side is continuous into $L^2$.)

An alternative approach to local existence is given in \cite{bbs}, where
it is shown that, in Coulomb gauge,
systems of the type \eqref{scs}
can be put in the form of an abstract evolution equation to which Kato's theory
(\cite{kato75}) applies. This yields the existence of a {\it local}
solution denoted $(A', \Phi ') $  with
$\Phi '$ continuous into $H^2$ on a time interval of length determined by the
$H^2$ norm of the initial data. But
the estimate \eqref{gig} above is gauge invariant, 
and allows continuation of
the local solution to provide a global solution in Coulomb gauge with
regularity $\Phi '\in  C\bigl([0,\infty); H^2 (\Sigma)\bigr)
\cap C^1\bigl([0,\infty); L^2 (\Sigma)\bigr)$
and $\mbA '\in  C\bigl([0,\infty); H^3 (\Sigma)\bigr)\cap
C^1\bigl([0,\infty); H^1(\Sigma)\bigr)$ satisfying the Coulomb gauge
condition  $\dv A' = 0$.

Finally, we explain how to obtain theorem \ref{thglob2} from these 
results.
Given a solution $\mbA',\Phi'$ in Coulomb gauge, 
as just described,
define $\chi(t,x)$ to be the solution of 
$$
(-\Delta+|\Phi'|^2)\dot\chi=\dv\dot \mbA'-\langle i\Phi',\dot\Phi'\rangle=-\langle i\Phi',\dot\Phi'\rangle,
$$
with $\chi(0,x)=0$.
Then it is easy to verify that 
$\mbA=\mbA'+\mathbf{d}\chi,\Phi=\Phi' e^{i\chi}$ satisfies
\eqref{gauge1}. Under the condition $\|\Phi(t)\|_{L^2(\Sigma)}^2=L>0$
the solution exists and is unique at time $t$; 
this condition is natural because
$\|\Phi(t)\|_{L^2(\Sigma)}$ is independent of time for solutions of
\eqref{scs}.
Now by the above mentioned Coulomb gauge regularity and the 
basic estimates 
for the Laplacian we deduce that $\chi\in C^1([0,\infty); H^2 )$. 
This 
gives the global existence theorem in the gauge stated in 
theorem \ref{thglob2}.


\small
\baselineskip=13pt
\begin{thebibliography}{10}


\bibitem{amt}
        {R. Abraham, J. Marsden and T. Ratiu},
        {\em Manifolds, tensor analysis and applications},
        {Springer verlag, New York, 1988}.

\bibitem{aswz}
{Arovas, D., Schrieffer, R., Wilczek, F. and Zee, A.}
{Statistical mechanics of anyons}
{\em Nucl. Phys. B} 
{\bf 251}, 117-126 (1985).


\bibitem{atz}
{Aitchison, I.J.R, Ao, P., Thouless, D. and Zhu, X}
{\em Phys. Rev B.} {\bf 51}  6531 (1995).



\bibitem{bl}
        {Y. Benyamini and J. Lindenstrauss},
        {\em Geometric nonlinear functional analysis},
        {American Mathematical Society, Providence, 2000}.

\bibitem{bbs}
         {L. Berge, A. de Bouard and J. Saut},
         {Blowing up time-dependent solutions of 
         the planar Chern-Simons gauged 
         nonlinear Schr\"odinger equation,}
         {\em Nonlinearity} {\bf 8} 235-253 (1995).



\bibitem{br95}
        {F. Bethuel and T. Riviere},
        Vortices for a variational problem related to
        superconductivity,
        {\em Ann. Inst. H. Poincar\'e  Anal. Non Lin\'eaire} 
        {\bf 12}(3) 243-303 (1995). 



\bibitem{bogo76}
                 {E. Bogomolny,}
                 Stability of Classical Solutions,
                 {\em Soviet Journal of Nulclear Physics}
                 {\bf 24} 861-870 (1976).




\bibitem{brad88}
         {S. Bradlow,}
         Vortices in holomorphic line bundles and closed Kaehler manifolds,
         {\em Commun. Math. Phys.} {\bf 118} 1-17 (1990).

\bibitem{braddask91}
          {S. Bradlow and G. Daskalopoulos},
          {Moduli of stable pairs for holomorphic bundles 
                     over Riemann surfaces},
          {\em Internat. J. Math.}
          {\bf 2} 477-513 (1991).

\bibitem{bg}
         {H. Brezis and T. Gallouet,}
         {\em Nonlinear analysis, theory methods and applications,}
         {\bf 4}, no. 4  677-681 (1980).




\bibitem{djt}
{Deser, S. and Jackiw, R. and Templeton, S.},
{Topologically massive gauge theories},
{\em Ann. Physics},
{\bf 140}, {372--411} (1982).

   
\bibitem{ds97}
       {S. Demoulini and D. Stuart},
       {Gradient flow of the superconducting Ginzburg-Landau
       functional on the plane},
       {\em Commun.  Anal. Geom.}
       {\bf 5}(1)
       {121~-~198 (1997).}






\bibitem{ds07}
       {S. Demoulini},
       {Global existence for a nonlinear Schr\"odinger-Chern-Simons system on a surface},
       {\em Ann. Inst. H. Poincar\'e Anal. Non Lin\'eaire}
       {\bf 24}(2)  207-225  (2007).

\bibitem{sdds07}
{S. Demoulini and D. M. A. Stuart}
{Existence and regularity for generalised harmonic maps 
associated to a nonlocal polyconvex energy of Skyrme type}
{\em Calculus of Variations and PDE}
{\bf 30}, 4 523-546 (2007).


\bibitem{frm}
{Froehlich, J.  \& Marchetti, P-A} 
{\em Comm. Math. Phys.}
{\bf 121}, 177-221 (1989).

\bibitem{fs92}
{Froehlich, J.  and Studer, U.M.} 
{\em $U(1)\times SU(2)$ - gauge invariance of non-relativistic 
quantum mechanics and generalized Hall effects.}
{\em Comm. Math. Phys.}
{\bf 148}, 553-600 (1992).
 





\bibitem{gd}
{Dunne, G.}
{Aspects of Chern-Simons theory},
{appearing in} {\em Les Houches lectures on 
Topological aspects of low dimensional systems},
{EDP Sci., Les Ulis}, {1998}. 
Available online at arXiv:hep-th/9902115v1


\bibitem{g99}
{Girvin, S.}
{The Quantum Hall Effect: Novel Excitations and Broken Symmetries},
{appearing in} {\em Les Houches lectures on 
Topological aspects of low dimensional systems},
{EDP Sci., Les Ulis}, {1998}.
Available online at arXiv:cond-mat/9907002v1


\bibitem{sg}{ S. Gustafson and I.M. Sigal},
            {The stability of magnetic vortices},
            {\em Commun. Math. Phys.}
            {\bf 212}
            {257--275 (2000).}

\bibitem{hs}
{Haskins, M. and Speight, J. M.} 
{The geodesic approximation for lump dynamics and coercivity of the
Hessian for harmonic maps,}
{J. Math. Phys.}  
{\bf 44}, 
{3470--3494 (2003).}




\bibitem{hh}
           {M. Hassaïne and P. Horvathy}, 
           {Non-relativistic Maxwell-Chern-Simons vortices},  
           {Ann. Physics}
           {\bf 263},  no. 2, 
           {276--294 (1998)}   

\bibitem{hz}
           {P. Horvathy and P. Zhang}, 
           {Vortices in abelian Chern-Simons gauge theory}.  
{Available online at arxiv:hep-th/0811.2094.}




\bibitem{jp}
{Jackiw, R. and Pi, So-Young},
{Self-dual Chern-Simons solitons},
{appearing in Low-dimensional field theories 
and condensed matter physics (Kyoto,1991)},
{\em Progr. Theoret. Phys. Suppl.},
{\bf 107}, {1-40} (1992).



\bibitem{jost98}
            {J. Jost},
            {\em Riemannian geometry and geometric analysis},
            {Springer-Verlag 1988.}  


\bibitem{kato80}
            {T. Kato},
            {\em Perturbation theory for linear operators,}
            {Springer-Verlag 1980.}  

\bibitem{kato75}
{T. Kato,}
{Quasi-linear equations of evolution with applications to partial differential equations},
{\em Springer Lecture Notes in Mathematics}
{\bf 448}  27--50 (1975).

\bibitem{ks}
{S. Krusch and P. Sutcliffe,}
{Schr\"odinger-Chern-Simons vortex dynamics,}  
{\em Nonlinearity}  
{\bf 19}  
{1515--1534 (2006)}



\bibitem{jt82} A. Jaffe and C. Taubes, 
               {\em Vortices and Monopoles}, 
               Birkhauser,  Boston, 1982.



\bibitem{mb}
{A. Majda and A. Bertozzi}, 
{\em Vorticity and Incompressible Fluid Flow},
{Cambridge University Press 2001.}

\bibitem{gobs97}
        N. Manton,
        First order vortex dynamics,
        {\em Ann. Phys.}
        {\bf 256} 114-131 (1997).
    
\bibitem{ms2004}
        N. Manton and P. Sutcliffe,
        {\em Topological Solitons},
        {Cambridge University Press 2004.}
        
    
\bibitem{n}
{Nagosa, N}
{\em Quantum Field Theory in Condensed Matter Physics}
{Springer, Berlin, 1999}


\bibitem{palais68}
            {R. Palais,}
            {Foundations of global nonlinear analysis,}
            {\em Mathematics lecture note series},
            {W.A. Benjamin, New York, 1968.}





\bibitem{pg}
{Prange, R. and Girvin, S.}
{\em The Quantum Hall effect}, 2nd edition
 {Springer verlag, New York, 1990}.





\bibitem{rsI}
            {M. Reed and B. Simon},
            {\em Functional analysis},
            {Academic press, San Diego 1980.}


\bibitem{rs}
{Rodnianski, I. and Sterbenz, J.}
{On the formation of singularities in the critical
$O(3)\,\sigma$-model. arXiv:math.AP/0605023.}


\bibitem{r}
{N. Romao}
{Quantum Chern-Simons vortices on a sphere},
{\em J. Math. Phys.}
{\bf 42}
{3445-3469 (2001)}

\bibitem{rt02}
{N. Romao}
{Classical and quantum aspects of topological solitons},
{PhD Thesis},
{University of Cambridge},
{(2002)}


\bibitem{rsp}
            {N. Romao and J.M. Speight},
            {Slow Schrödinger dynamics of gauged vortices},  
            {\em Nonlinearity}  
            {\bf 17} no. 4, 
            {1337--1355 (2004)}

\bibitem{rh57}
            {H. Rubin and P. Ungar},
            {Motion under a strong constraining force},
            {\em Commun. Pure Appl. Math.}
            {\bf 10}            
            {65-87 (1957).}

\bibitem{ss07}
            {E. Sandier and S. Serfaty},
            {Vortices in the Magnetic Ginzburg-Landau Model},
            {\em Progress in Nonlinear Differential Equations 
and their Applications}
            {\bf 70} 
            Birkhauser, (2007). 




\bibitem{skkr}        
{S. L. Sondhi, A. Karlhede, S. A. Kivelson, and E. H. Rezayi},
{Skyrmions and the crossover from the integer to fractional 
quantum Hall effect at small Zeeman energies},
{\em Phys. Rev. B} 
{\bf 47} 16419 (1993).


\bibitem{sto90}
{Stone, M}
{Superfluid dynamics of the fractional quantum Hall state}
{\em Physical Review B}
{\bf 42}, 1 212 (1990).


\bibitem{sto}
{Stone, M}
{\em Int. J. Mod. Phys. B} {\bf 9}, 
1359 (1995).





\bibitem{s94}
       {D. Stuart},
       {Dynamics of Abelian Higgs vortices in the near Bogomolny
       regime},
       {\em Commun. Math. Phys.}
       {\bf 159}
       {51-91 (1994).}


\bibitem{s94b}
       {D. Stuart},
       {The geodesic approximation for the 
        Yang-Mills-Higgs equations},
       {\em Commun. Math. Phys.}
       {\bf 166}
       {149-190 (1994).}



\bibitem{s99}
        {D. Stuart},
        {Periodic solutions of the Abelian Higgs model and rigid
        rotation of vortices},
        {\em Geom. Funct. Anal.}
        {\bf 9} 1-28 (1999).

\bibitem{s07}
        {D. Stuart},
        {Analysis of the adiabatic limit for solitons in classical field theory},
        {\em Proc R Soc A}
        {\bf 463} 2753-2781 (2007).

\bibitem{taylor}
         {M. Taylor},
         {\em Partial Differential Equations},
{Applied Mathematical Sciences, vol 117},         
{Springer-Verlag 1996.}


\bibitem{ts03}
{A. M. Tsvelik},
{\em Quantum Field Theory in Condensed Matter Physics},
{Cambridge University Press},
{Cambridge 2003.}

	
\bibitem{wilc82}
{F. Wilczek}
{Quantum mechanics of fractional spin particles}
{\em Physical Review Letters}
{\bf 49}, 1 957 (1982).    

\bibitem{zhang92}
{Zhang, S.C.}
{The Chern-Simons-Landau-Ginzburg theory of the fractional 
quantum Hall effect}
{\em Int. J. Mod. Phys. B}
{\bf 6}, 1  43-77 (1992).


\bibitem{zhk}
{Zhang, S.C., Hansson, T.H. and Kivelson, S.}
{Effective field theory model for the fractional quantum Hall effect},
{\em Phys. Rev. Lett.} {\bf 62}  82 (1989). 
Erratum: {\em Phys. Rev. Lett.} {\bf 62}  980 (1989). Available online
at http://prola.aps.org.




\end{thebibliography}
\end{document}